%% file: eetojpsidd.latex
\documentclass[aps,prd,twocolumn,superscriptaddress,showpacs,preprintnumbers,notitlepage]{revtex4-1}

\usepackage{graphicx}
\usepackage{tensor}
\usepackage{multirow}
\usepackage{xcolor}
\usepackage{amsmath}
\usepackage{mathtools}
\graphicspath{{images/}}

\newcommand{\err}[2]{\tensor*[^{+#1}_{-#2}]{}{}}

\newcommand{\elp}{{e^+}}
\newcommand{\elm}{{e^-}}

\newcommand{\mup}{{\mu^+}}
\newcommand{\mum}{{\mu^-}}

\newcommand{\lp}{{\ell^+}}
\newcommand{\lm}{{\ell^-}}
\newcommand{\leplep}{{\ell \ell}}
\newcommand{\pip}{{\pi^+}}
\newcommand{\pim}{{\pi^-}}

\newcommand{\piz}{{\pi^0}}
\newcommand{\kp}{{K^+}}
\newcommand{\km}{{K^-}}

\newcommand{\ks}{{K^0_S}}
\newcommand{\D}{{D}}
\newcommand{\Db}{{\bar{D}}}
\newcommand{\Dp}{{D^+}}

\newcommand{\Dz}{{D^0}}

\newcommand{\DD}{{\D \Db}}
\newcommand{\jp}{{J/\psi}}

\newcommand{\jpdd}{{\elp \elm \to \jp \DD}}
\newcommand{\gisr}{{\gamma_\mathrm{ISR}}}

\newcommand{\chicp}{{\chi_{c0}(2P)}}

\newcommand{\x}[1]{{X(#1)}}
\newcommand{\xst}[1]{{X^*(#1)}}
\newcommand{\Xst}{{X^*}}
\newcommand{\y}[1]{{Y(#1)}}

\newcommand{\mrjpd}{M_{\mathrm{rec}}}
\newcommand{\thprod}{\theta_\mathrm{prod}}
\newcommand{\lbeam}{\lambda_\mathrm{beam}}
\newcommand{\Dfun}[6]{D^{#1}_{#2\,#3}(#4, #5, #6)}
\newcommand{\dfun}[4]{d^{#1}_{#2\,#3}(#4)}

\newcommand{\fb}{\mathrm{fb}}
\newcommand{\invfb}{\mathrm{fb}^{-1}}
\newcommand{\cm}{\mathrm{cm}}
\newcommand{\mev}{\mathrm{MeV}}
\newcommand{\mevcc}{\mathrm{MeV}/c^2}
\newcommand{\gev}{\mathrm{GeV}}

\newcommand{\gevcc}{\mathrm{GeV}/c^2}

\newcommand{\dcha}[1]{
\ifcase#1\relax
\Dp \to \ks \pip
\or
\Dp \to \km \pip \pip
\or
\Dp \to \ks \pip \piz
\or
\Dp \to \km \pip \pip \piz
\or
\Dp \to \ks \pip \pip \pim
\or
\Dp \to \ks \ks \kp
\or
\Dp \to \ks \kp
\or
\Dz \to \km \pip
\or
\Dz \to \ks \piz
\or
\Dz \to \ks \pip \pim
\or
\Dz \to \km \pip \piz
\or
\Dz \to \km \pip \pip \pim
\or
\Dz \to \ks \pip \pim \piz
\or
\Dz \to \ks \kp \km
\or
\Dz \to \kp \km
\fi
}

\newcommand{\dlnl}{\Delta(-2 \ln L)}
\newcommand{\br}{\mathcal{B}}
\newcommand{\babar}{{\it BABAR }}

\begin{document}

\input{authors.tex}

\title{Observation of an alternative $\chicp$ candidate in $\jpdd$}


\begin{abstract}
We perform a full amplitude analysis of the process $\jpdd$,
where $\D$ refers to either $\Dz$ or $\Dp$.
A new charmoniumlike state $\xst{3860}$ that decays to $\DD$
is observed with a significance of $6.5\sigma$.
Its mass is $(3862\err{26}{32}\err{40}{13})\ \mevcc$ and width is
$(201\err{154}{67}\err{88}{82})\ \mev$.
The $J^{PC}=0^{++}$ hypothesis is favored over the $2^{++}$ hypothesis at the
level of $2.5\sigma$.
The analysis is based on the 980 $\invfb$ data sample collected by the Belle
detector at the asymmetric-energy $\elp \elm$ collider KEKB.
\end{abstract}




\maketitle

\section{Introduction}

The charmoniumlike state $\x{3915}$ was observed by the Belle
Collaboration in $B \to \jp \omega K$ decays~\cite{Abe:2004zs};
its original name was the $\y{3940}$.
Subsequently, it was also observed by the \babar Collaboration in the same
$B$ decay mode~\cite{Aubert:2007vj,delAmoSanchez:2010jr} and by both
Belle~\cite{Uehara:2009tx} and \babar~\cite{Lees:2012xs} in
the process $\gamma\gamma \to \x{3915} \to \jp\omega$.
The quantum numbers of the $\x{3915}$ were
measured in Ref.~\cite{Lees:2012xs} to be $J^{PC} = 0^{++}$.
As a result, the $\x{3915}$ was identified as the
$\chicp$ in the 2014 PDG tables~\cite{Agashe:2014kda}.

However, this identification remains in doubt.
The $\chicp$ is expected to decay strongly to $\DD$ in an $S$-wave.
Here and elsewhere, $\D$ refers to either $\Dz$ or $\Dp$. The
$\chicp \to \DD$ decay mode is expected to be dominant,
but has not yet been observed experimentally for the $\x{3915}$;
in contrast, the decay mode that is observed, $X(3915) \to \jp \omega$,
would be OZI (Okubo-Zweig-Iizuka)~\cite{ozi} suppressed
for the $\chicp$. Since the $\chicp$ should decay to $\DD$ in an $S$-wave,
this state is na\"{\i}vely expected to have a large width of
$\Gamma \gtrsim 100\ \mev$~\cite{Guo:2012tv}; however, the
measured $\x{3915}$ width of $(20 \pm 5)\ \mev$ is much
smaller. We note that there are some specific predictions for the $\chicp$
width that give small values.
For example, the predicted width is $30\ \mev$ in Ref.~\cite{Barnes:2005pb}
and between $12$ and ${\sim}100\ \mev$, depending on the $\chicp$ mass, in
Ref.~\cite{Eichten:2004uh}.
If the $\chicp$ and $\x{3915}$ are the same state, then
the partial width to $\jp \omega$, which has been
estimated to be
$\Gamma(\chicp \to \jp \omega) \gtrsim 1\ \mev$~\cite{Guo:2012tv}, is too
large.
Also, in this case,
one can obtain contradictory estimates for the branching fraction
$\br(\chicp \to \jp \omega)$~\cite{Olsen:2014maa}.
Furthermore, the $\chi_{c2}(2P)$ $-$ $\chicp$ mass splitting
would be much smaller than the prediction of potential
models and smaller than the mass difference for
the bottomonium states $\chi_{bJ}(2P)$~\cite{Guo:2012tv,Olsen:2014maa},
which is 
inconsistent with expectations based on the heavier bottom quark mass.

The quantum numbers of the $\x{3915}$ were
measured in Ref.~\cite{Lees:2012xs} to be $J^{PC} = 0^{++}$
assuming that the $X(3915)$ is produced in
the process $\gamma\gamma\to X(3915)$ with helicity $\lambda=2$ if its
quantum numbers are $J^{PC}=2^{++}$.
A reanalysis of the data from Ref.~\cite{Lees:2012xs},
presented in Ref.~\cite{Zhou:2015uva}, shows that the $2^{++}$ assignment
becomes possible if the $\lambda = 2$ assumption is abandoned. As a result of
these considerations, the
$X(3915)$ is no longer identified as the $\chicp$ in the 2016 PDG
tables~\cite{Olive:2016xmw}.

It is possible that an alternative $\chicp$ candidate was actually observed
by Belle~\cite{Uehara:2005qd} and \babar~\cite{Aubert:2010ab}
in the process
$\gamma\gamma \to \DD$ together with the $\chi_{c2}(2P)$.
An alternative fit to Belle and
\babar data was performed in Ref.~\cite{Guo:2012tv}. In this fit,
the nonresonant $\gamma\gamma \to \DD$ events are attributed
to a wide charmonium state with mass and width
$M = (3838 \pm 12)\ \mevcc$ and $\Gamma = (211 \pm 19)\ \mev$.
However, in this estimate, additional signal sources such as feed-down
from $\gamma\gamma \to D\bar{D}^*$ events were not
taken into account;
consequently, if the $\chicp$ is really observed in the process
$\gamma\gamma \to \DD$, then the parameters measured in Ref.~\cite{Guo:2012tv}
are biased --- the $\chicp$ mass is shifted to lower
values~\cite{Olsen:2014maa}.

It is also possible to search for the $\chicp$ produced in the
process $\gamma\gamma\to\chicp$ using final states other than $\DD$ and
$\jp\omega$. Belle searched for
high-mass charmonium states in the process
$\gamma\gamma\to\ks\ks$~\cite{Uehara:2013mbo}.
An excess with a marginal statistical significance of $2.6\sigma$ was
found in the mass region between 3.80 and 3.95 $\gevcc$.

A unique process that is suitable for
a search for the $\chicp$ and
other charmonium states with positive $C$-parity is double-charmonium
production in association with the $\jp$.
The $C$-parity of the states observed with the $\jp$
is guaranteed to be positive in the case of one-photon annihilation; the
annihilation via two virtual photons is suppressed by a factor
$(\alpha / \alpha_s)^2$~\cite{Bodwin:2002kk,Bodwin:2002fk}.
The charmoniumlike $\x{3940}$ state was observed by Belle in the inclusive
$\elp \elm \to \jp X$ spectrum and in
the process $\elp \elm \to \jp D^* \bar{D}$~\cite{Abe:2007jna,Abe:2007sya},
and the $\x{4160}$ was observed in the process
$\elp \elm \to \jp D^* \bar{D}^*$~\cite{Abe:2007sya}.
Indications of a $\DD$ resonance in $\elp \elm \to \jp D \bar{D}$
were reported in Ref.~\cite{Abe:2007sya}; however, the existence of
this resonance could not be established reliably.

Here, we present an updated analysis of the process $\jpdd$.
The analysis is performed using the $980\ \invfb$ data sample
collected by the Belle detector
at the asymmetric-energy $\elp \elm$ collider KEKB~\cite{kekb}.
The data sample was collected at or near the
$\Upsilon(1S)$, $\Upsilon(2S)$, $\Upsilon(3S)$, $\Upsilon(4S)$ and
$\Upsilon(5S)$ resonances. The integrated luminosity is 1.4 times greater than
the luminosity used in the previous analysis~\cite{Abe:2007sya}.
In addition, we use a multivariate method to improve the discrimination
of the signal and background events and an amplitude analysis
to study the $J^{PC}$ quantum numbers of the $\DD$ system.

\section{The Belle detector}

The Belle detector is a large-solid-angle magnetic
spectrometer that consists of a silicon vertex detector (SVD),
a 50-layer central drift chamber (CDC), an array of
aerogel threshold Cherenkov counters (ACC),
a barrel-like arrangement of time-of-flight
scintillation counters (TOF), and an electromagnetic calorimeter
comprised of CsI(Tl) crystals (ECL) located inside
a superconducting solenoid coil that provides a 1.5~T
magnetic field.  An iron flux-return located outside of
the coil is instrumented to detect $K_L^0$ mesons and to identify
muons (KLM).  The detector
is described in detail elsewhere~\cite{Belle}.
Two inner detector configurations were used. A 2.0 cm beampipe
and a three-layer silicon vertex detector were used for the first sample
of 156 $\invfb$, while a 1.5 cm beampipe, a four-layer
silicon detector and a small-cell inner drift chamber were used to record
the remaining data~\cite{svd2}.

We use a GEANT-based Monte Carlo (MC) simulation~\cite{geant} to model
the response of the detector, identify potential backgrounds and
determine the acceptance. The MC simulation includes run-dependent
detector performance variations and background conditions. The signal MC
generation is described in Sec.~\ref{sec:mva}.

\section{Reconstruction}
\label{sec:reconstruction}

We select events of the type $\jpdd$,
where only the $\jp$ and one of the $\D$ mesons
are reconstructed; the other $\D$ meson is identified
by the recoil mass of the $(\jp,\D)$ system, which is denoted as $\mrjpd$.

All tracks are required to originate from the interaction point region:
we require $dr < 0.2\ \cm$ and $|dz| < 2\ \cm$, where
$dr$ and $dz$ are the cylindrical coordinates of the point of the
closest approach of the track to the beam axis
(the $z$ axis of the laboratory reference frame coincides with
the positron-beam axis).

Charged $\pi$ and $K$ mesons are identified using
likelihood ratios $R_{\pi/K} = \mathcal{L}_\pi/(\mathcal{L}_\pi+\mathcal{L}_K)$
and $R_{K/\pi} = \mathcal{L}_K/(\mathcal{L}_\pi+\mathcal{L}_K)$,
where $\mathcal{L}_\pi$ and $\mathcal{L}_K$ are the likelihoods for
$\pi$ and $K$, respectively. The likelihoods
are calculated from the combined
time-of-flight information from the TOF, the number of photoelectrons from
the ACC and $dE/dx$ measurements in the CDC.
We require $R_{K/\pi} > 0.6$ for $K$ candidates and $R_{\pi/K} > 0.1$ for
$\pi$ candidates. The $K$ ($\pi$) identification efficiency is
typically 90\% (98\%) and the misidentification probability for non-$K$
(-$\pi$) background is about 4\% (30\%).

Electron candidates are identified as CDC charged tracks
that are matched to electromagnetic showers in the ECL.
The track and ECL cluster matching quality,
the ratio of the electromagnetic shower energy to the track momentum,
the transverse shape of the shower, the ACC light yield and the 
track's $dE/dx$ ionization
are used in our electron identification selection criteria.
A similar likelihood ratio is constructed:
$R_e = \mathcal{L}_e / (\mathcal{L}_e + \mathcal{L}_h)$,
where $\mathcal{L}_e$ and $\mathcal{L}_h$ are the likelihoods for electrons and
charged hadrons ($\pi$, $K$ and $p$), respectively~\cite{Hanagaki:2001fz}.
An electron veto ($R_e < 0.9$) is imposed on $\pi$ and $K$ candidates.
The veto rejects from 4 to 11\% of background events, depending on
the $D$ decay channel, while its signal efficiency is more than 99\%.

Muons are identified by their range and transverse scattering in the KLM.
The muon likelihood ratio is defined as
$R_\mu = \mathcal{L}_\mu / (\mathcal{L}_\mu + \mathcal{L}_\pi + \mathcal{L}_K)$,
where $\mathcal{L}_\mu$ is the likelihood for muons~\cite{Abashian:2002bd}.

The $\piz$ candidates are reconstructed via their
decay to two photons; we require
their energies to be greater than $30\ \mev$ and their invariant mass
$M_{\gamma\gamma}$ to satisfy $|M_{\gamma\gamma} - m_{\piz}| < 15\ \mevcc$,
where $m_{\piz}$ is the nominal $\piz$ mass~\cite{Agashe:2014kda}.
This requirement corresponds approximately
to a $3\sigma$ mass window around the nominal mass.

Candidate $\ks$ mesons are reconstructed from pairs of oppositely charged
tracks (assumed to be pions) and selected
on the basis of the $\pip \pim$ invariant mass,
the candidate $\ks$ momentum and decay angle,
the inferred $\ks$ flight distance in the $xy$ plane,
the angle between the $\ks$ momentum and the direction from the
interaction point to the $\ks$ vertex,
the shortest $z$ distance between the two pion tracks,
their radial impact parameters and numbers of hits in the SVD and CDC.

The $\jp$ is reconstructed via its $\elp \elm$ and $\mup \mum$ decay channels.
For $\jp \to \elp \elm$ candidates, we include photons that have energies
greater than 30 $\mev$ and are within 50 mrad of
a daughter lepton candidate momentum direction
in the calculation of the $\jp$ invariant mass. The $\jp$ daughter leptons
are identified as electrons or muons ($R_e > 0.1$ or $R_\mu > 0.1$) and
not as kaons ($R_{K/e} < 0.9$ and $R_{K/\mu} < 0.9$ for
the $\jp \to \elp \elm$ and $\jp \to \mup \mum$ decay modes, respectively).
The combined lepton-pair identification efficiency
is 91\% and 75\% and the fake rate is 0.07\% and 0.13\% for the
$\jp\to\elp \elm$ and $\jp \to \mup \mum$ channel, respectively.
We retain candidates that satisfy $|M_\jp - m_\jp| < 300\ \mevcc$, where
$M_\jp$ is the reconstructed mass and $m_\jp$ is the nominal
mass~\cite{Agashe:2014kda}. The mass window is intentionally
wide because this selection is at a preliminary stage and
includes the events that will be
used for the background determination.

The $\Dp$ mesons are reconstructed in five decay
channels: $\ks \pip$, $\km \pip \pip$, $\ks \pip \piz$,
$\km \pip \pip \piz$, and $\ks \pip \pip \pim$.
The $\Dz$ mesons are reconstructed in four decay channels: $\km \pip$,
$\ks \pip \pim$, $\km \pip \piz$, and $\km \pip \pip \pim$.
We retain candidates that satisfy $-150\ \mevcc < M_\D - m_\D < 350\ \mevcc$,
where $M_\D$ is the reconstructed mass and $m_\D$ is the nominal
mass~\cite{Agashe:2014kda}. The mass window is chosen to be
asymmetric to exclude peaking backgrounds from partially reconstructed
multibody $D$ decays, for example $\dcha{10}$ or $\dcha{1}$ backgrounds
in the $\dcha{7}$ event sample.

After the selection of the $\jp$ and $\D$ candidates, we
perform a mass-constrained fit to each candidate; then, the recoil mass
$\mrjpd$ is calculated. We retain the ($\jp$, $\D$) pairs that satisfy
$|\mrjpd - m_\D| < 350\ \mevcc$. Finally, we perform a beam-constrained
fit with $\mrjpd$ constrained
to the $\D$ mass to improve the $M_\DD$ resolution.

\section{Multivariate analysis and global optimization}
\label{sec:mva}

\subsection{Multivariate analysis}

To improve the separation of signal and background events,
we perform a multivariate analysis for each individual $D$ decay channel
followed by a global selection requirement optimization.
For each $D$ channel, the global
optimization includes the definition of the signal region and the requirement
on the multivariate-analysis output.

The algorithm used for the multivariate analysis is the multilayer
perceptron
(MLP) neural network implemented in the TMVA library~\cite{tmva}.
For $\D$ decay channels without a daughter $\piz$, the neural network contains
five variables: the cylindrical coordinate $dr$ of the $\jp$ vertex,
the cylindrical coordinates $dr$ and $dz$ of the $\D$ vertex calculated
relative to the $\jp$ vertex (these coordinates being sensitive to the $\D$
flight distance), the angle between the $\D$
momentum and the direction from the $\jp$ vertex to the $\D$ vertex,
and
the smaller of the two lepton-likelihood ratios $R_e$ or $R_\mu$ of the
$\jp$ daughter leptons for the $\jp\to\elp \elm$ and $\jp \to \mup \mum$
channel, respectively. If the $\D$ meson has a daughter $\piz$, three
additional $\piz$-related variables are used in the neural network:
the minimum energy of the $\piz$ daughter photons in the
laboratory reference
frame, the $\piz$ mass, and the angle between $-\vec{p}_\D$ and
$\vec{p}_\gamma$,
where $\vec{p}_\D$ and $\vec{p}_\gamma$ are the momenta of the parent
$\D$ and daughter $\gamma$ in the $\piz$ rest frame.

The signal data sample consists of Monte-Carlo events generated by a
combination of the PHOKHARA 9~\cite{phokhara9} and EvtGen~\cite{evtgen}
generators. The PHOKHARA 9 generator generates the initial-state radiation
photon(s); the remaining generation is performed by EvtGen.
The background sample is taken from the data; a two-dimensional ($M_\jp, M_\D$)
sideband is used. The background region includes all initially selected events
except the events with $|M_\jp - m_\jp| < 100\ \mevcc$ and
$|M_\D - m_\D| < 50\ \mevcc$.
In addition, we select only the events with the $\DD$ invariant
mass $M_\DD < 6\ \gevcc$ for both signal and background samples. This
requirement improves the signal and background separation, because the angle
between the $\D$ momentum and the direction from the $\jp$ vertex
to the $\D$ vertex provides more effective separation for $\D$ mesons with
larger momentum.
The background data sample is split randomly
into training and testing parts of equal size.

\begin{figure}
\includegraphics[width=8cm]{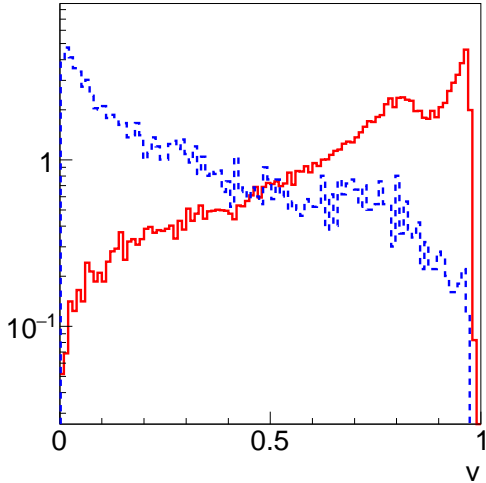}
\caption{Example of the MLP output (for the channel $\dcha{10}$). The red solid
line is the MLP output for the signal events and the blue
dashed line is the
MLP output for the background events. Both distributions are normalized so that
their integrals are equal to one.}
\label{fig:mlp_output}
\end{figure}

Example distributions of the MLP output $v$ for the signal and
background samples are shown in Fig.~\ref{fig:mlp_output}. The signal events
tend to have a larger value of $v$, while the background events have
smaller values of $v$. Thus, the requirements
on the MLP output used in the global selection requirement optimization have
the form $v > v_0$, where $v_0$ is a cutoff value.

\subsection{Fit to the background}

\begin{figure*}
\includegraphics[width=5.5cm]{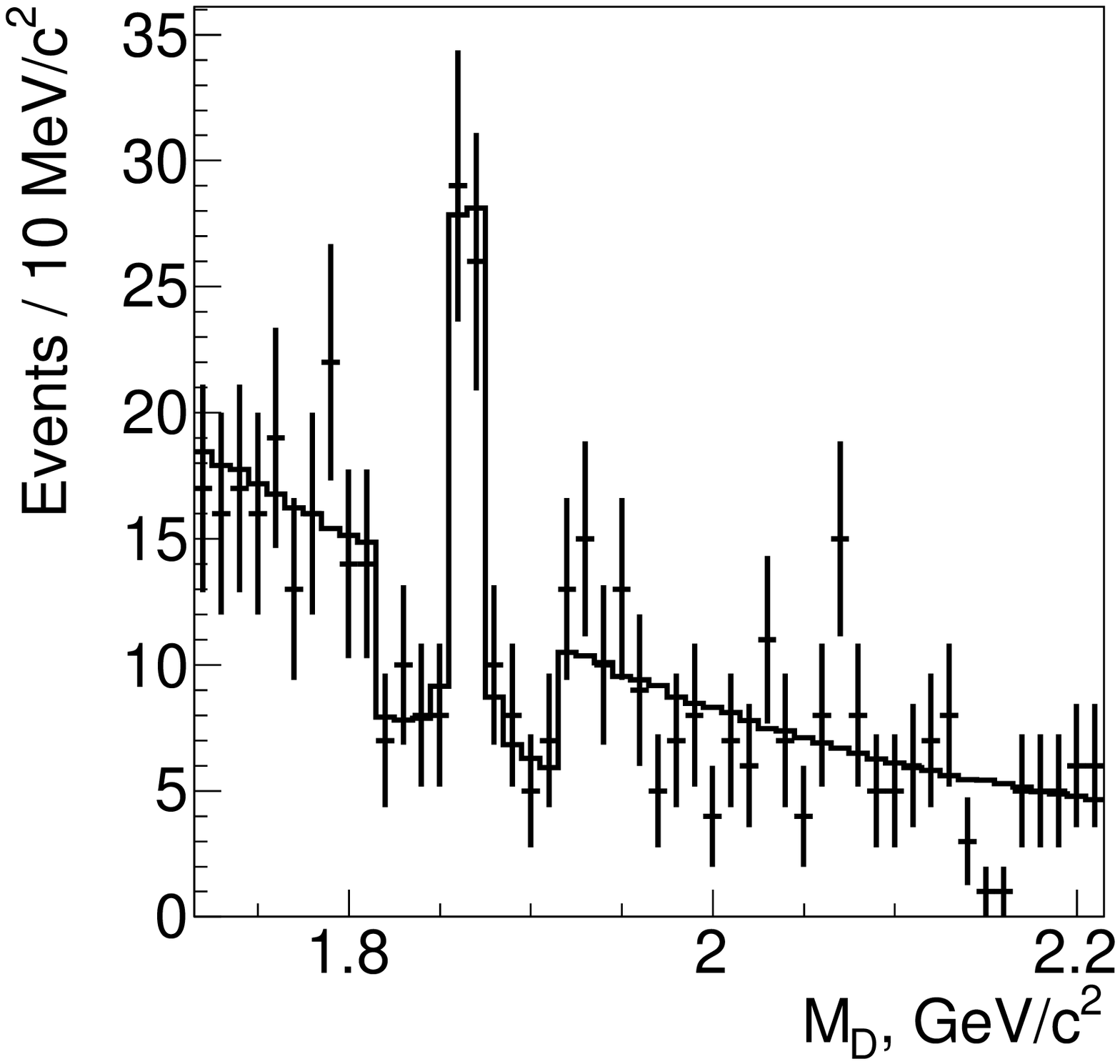}
\includegraphics[width=5.5cm]{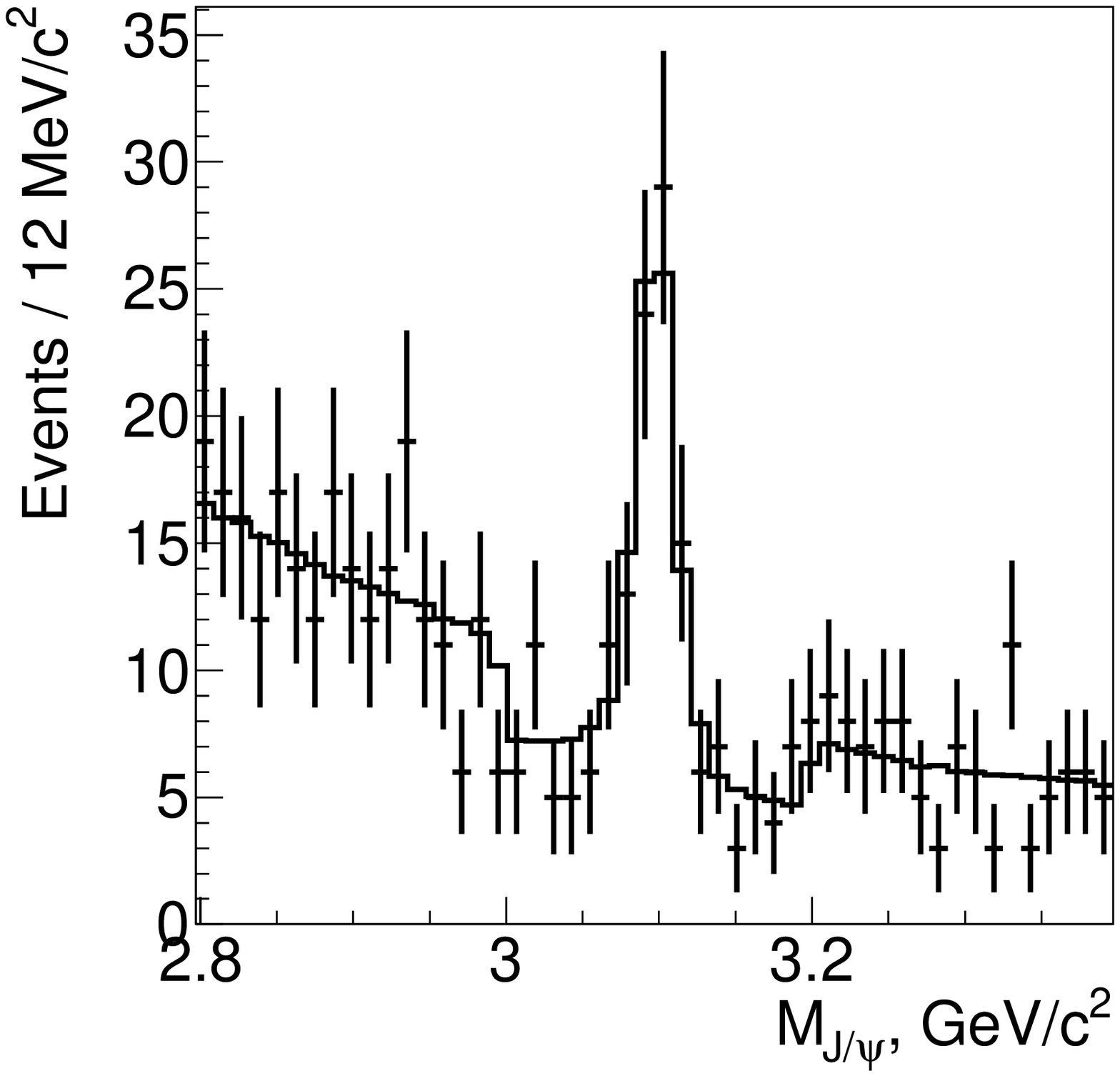}
\includegraphics[width=5.5cm]{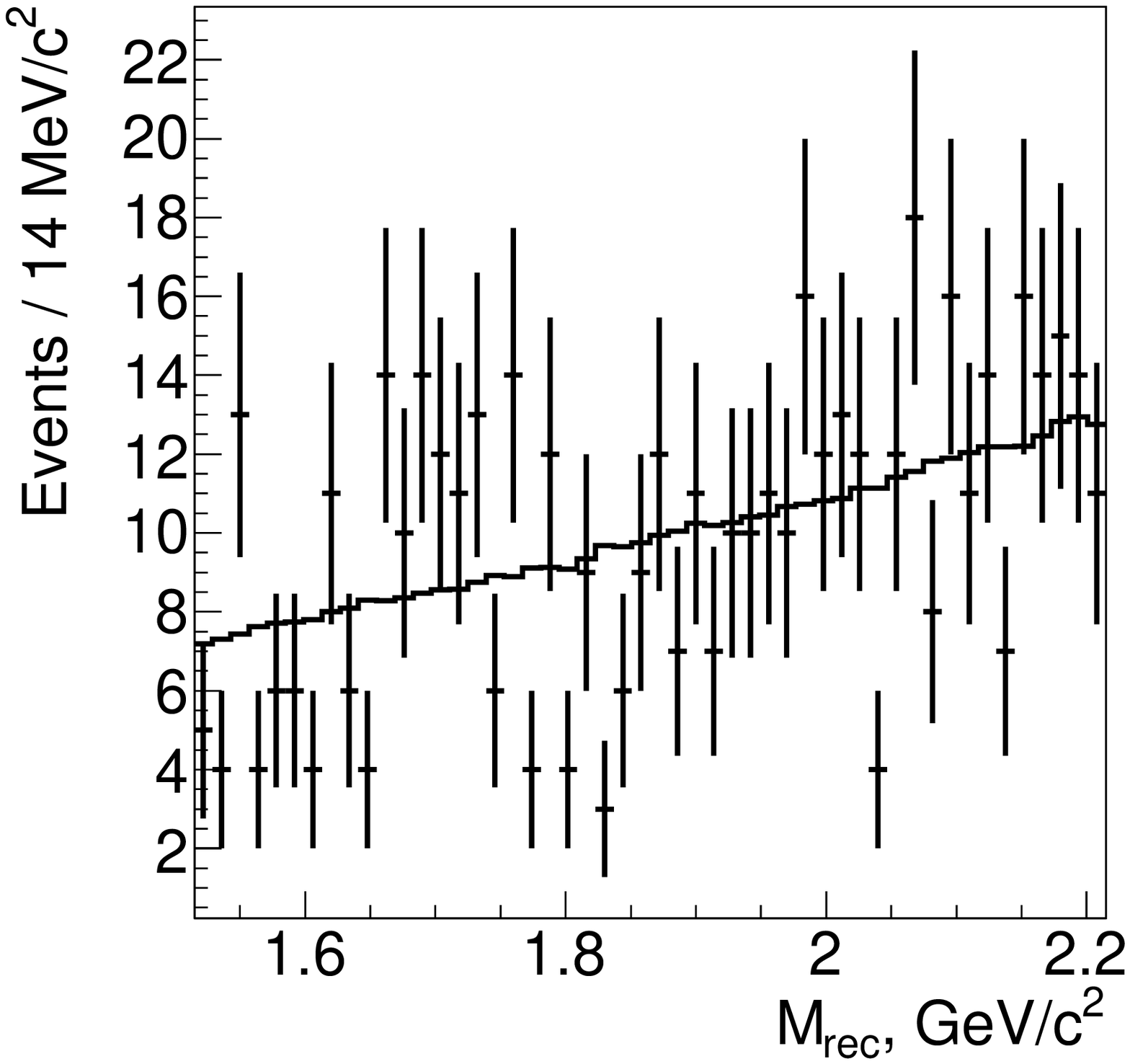}
\caption{Example of the results of the fit to the background events (for the
channel $\dcha{7}$). The points with error bars are the data and
the solid line is the fit result. The region defined by
$|M_\jp - m_\jp| < 100\ \mevcc$, $|M_\D - m_\D| < 50\ \mevcc$
[the region labeled (4) in Fig.~\ref{fig:background_subregions}] is excluded
from the fit; because of this exclusion, the projections of both the data
and the fit result onto $M_\D$ and $M_\jp$ have jump discontinuities.}
\label{fig:background_fit}
\end{figure*}

Before the global optimization, the events in the background region are
fitted to estimate the expected number of the background events in
the signal region. An unbinned maximum likelihood fit in the three-dimensional
parameter space $\Phi_\text{B} = (M_\jp, M_\D, \mrjpd)$ is performed.
The background density function is
\begin{equation}
\begin{aligned}
B(\Phi_\text{B}) = & P_2(\Phi_\text{B}) \exp(-\alpha M_\jp) \\
          & + R_{M_\D}(M_\D) P_1^{D}(M_\jp, \mrjpd) \\
          & + R_{M_\jp}(\tilde{M}_\jp) P_1^{\jp}(M_\D, \mrjpd), \\
\end{aligned}
\label{eq:bgpdf}
\end{equation}
where $P_2$ is a three-dimensional second-order polynomial, $P_1^{\D}$
and $P_1^\jp$ are two-dimensional first-order polynomials, $\alpha$ is a
rate parameter, $R_{M_\D}$ and $R_{M_\jp}$ are the resolutions in $M_\D$
and $M_\jp$, respectively, and $\tilde{M}_\jp$ is the scaled $\jp$ mass:
\begin{equation}
\tilde{M}_\jp = m_\jp + \mathcal{S} (M_\jp - m_\jp),
\label{eq:mass_scaling}
\end{equation}
where $\mathcal{S}$ is the scaling coefficient. This scaling
accounts for the fact that
the MC resolution in the $\jp$ mass is significantly better than
that measured with data.
The resolutions are determined from a fit of the distributions
of $M_\jp$ and $M_\D$ in the signal MC to a sum of an asymmetric
Gaussian and asymmetric
double-sided Crystal Ball functions~\cite{skwarnicki}. Some of the coefficients
of the polynomials are fixed at 0 for the $D$ channels that have a small number
of the background events.
The average value of the $\jp$ mass resolution scaling coefficient for all $D$
channels is measured to be $\mathcal{S} = 0.79 \pm 0.02$.
If the $\D$ mass resolution is scaled similarly to Eq.~\eqref{eq:mass_scaling},
the scaling coefficient depends on the
$\D$ decay channel; it is possible to determine this coefficient from data
only for channels with a large number of real $\D$ mesons in the background
region:
$\dcha{1}$ and $\dcha{7}$. The resolutions in data and MC are found to be
consistent for both $\dcha{1}$ and $\dcha{7}$; thus, the resolution for the
$\D$ mesons is not scaled.
An example of the background fit results
(for the channel $\dcha{7}$) is shown in Fig.~\ref{fig:background_fit}.
The fit quality is estimated using the mixed-sample
method~\cite{Williams:2010vh}; the confidence levels are found to be not
less than $15\%$. 

\begin{figure}
\includegraphics[width=8cm]{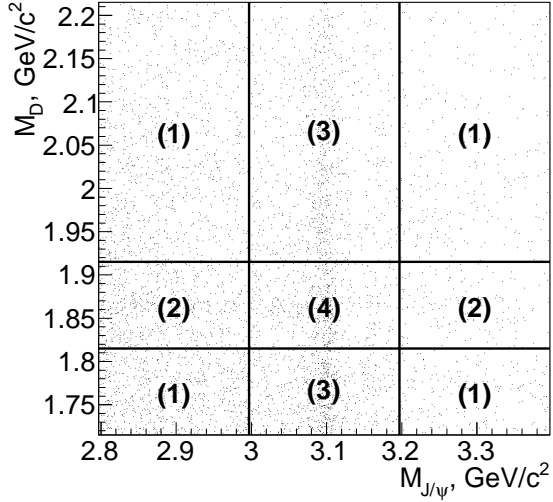}
\caption{Definition of the background subregions: (1) is the background region
where both $\D$ and $\jp$ mesons are fake, (2) is the real $\D$ region,
(3) is the real $\jp$ region. The central rectangle (4) is excluded from the
background region as described in Sec.~\ref{sec:reconstruction}.}
\label{fig:background_subregions}
\end{figure}

\subsection{Global selection requirement optimization}
\label{sec:selection_optimization}

The global selection requirement optimization is performed by maximizing the
value
\begin{equation}
\frac{\sum\limits_i N_\mathrm{sig}^{(i)}}
{\displaystyle{\frac{a}{2}} + \sqrt{\sum\limits_i N_\text{bg}^{(i)}}},
\label{eq:optfun}
\end{equation}
where $i$ is the index of the $\D$ decay channel, $N_\mathrm{sig}^{(i)}$
is the expected number of the signal events for the $i$-th channel,
$N_\text{bg}^{(i)}$ is the expected number of the background events
in the signal region,
and $a=5$ is the target significance.
This optimization method is based on Ref.~\cite{Punzi:2003bu}.
The signal region is defined as
\begin{equation}
\Big(\frac{M_\jp-m_\jp}{M_\jp^{(i)}}\Big)^2 +
\Big(\frac{M_\D-m_\D}{M_\D^{(i)}}\Big)^2 < 1,
\label{eq:sigreg_jpd}
\end{equation}
\begin{equation}
\Big|\frac{\mrjpd-m_\D}{\mrjpd^{(i)}}\Big| < 1, \\
\label{eq:sigreg_mrec}
\end{equation}
where $M_\jp^{(i)}$ and $M_\D^{(i)}$ are the half-axes of the
signal region ellipse and
$\mrjpd^{(i)}$ is the half-width of the selected recoil mass region.
All the values $M_\jp^{(i)}$, $M_\D^{(i)}$ and $\mrjpd^{(i)}$ are
channel-dependent.

The expected number of signal events is given by
\begin{equation}
\begin{aligned}
N_\mathrm{sig}^{(i)} = &
\sigma_{\jpdd}^\mathrm{(E_\gamma < E_\mathrm{cut})}(\Upsilon(4S))
f^{(i)}
\frac{N_{\text{rec}}^{(i)}}{N_{(\Upsilon(4S)\ E_\gamma < E_\mathrm{cut})}^{(i)}} \\
&\times
\mathcal{L}_{\Upsilon(4S)}
\epsilon_S^{(i)}(v_0^{(i)})
\int\limits_{\mathrm{SR}} S^{(i)}(\Phi) \,d \Phi, \\
\end{aligned}
\label{eq:nsig}
\end{equation}
where
\begin{equation}
\sigma_{\jpdd}^\mathrm{(E_\gamma < E_\mathrm{cut})} =
\sigma_{\jpdd}^\mathrm{(Born)}
\frac
{\sigma_{\elp \elm \to \mup \mum}^{(E_\gamma < E_\mathrm{cut})}}
{\sigma_{\elp \elm \to \mup \mum}^\mathrm{(Born)}} \\
\label{eq:cross_section_cut}
\end{equation}
is the cross section of the process $\jpdd(\gisr)(\gisr)$ with the
condition that the photon(s) energy is less than the cutoff energy and
the vacuum polarization taken into account,
$\sigma_\text{process}^\mathrm{(Born)}$
is the Born cross section of the specified process,
$f^{(i)}$ is the fraction of $i$-th channel,
$N_{(\Upsilon(4S)\ E_\gamma < E_\mathrm{cut})}^{(i)}$ is the
number of generated MC events at the $\Upsilon(4S)$ resonance with
$E_\gamma < E_\mathrm{cut}$,
$N_{\text{rec}}^{(i)}$ is the total number of reconstructed MC events
for all beam energies,
$\mathcal{L}_{\Upsilon(4S)}$
is the integrated luminosity at the $\Upsilon(4S)$ resonance,
$\epsilon_S^{(i)}(v_0^{(i)})$ is the efficiency of the
requirement ($v > v_0^{(i)}$) on the MLP output $v$ for the signal events,
SR is the signal region, and
$S^{(i)}(\Phi)$ is the signal probability density function (PDF),
which is proportional to the
product of the resolutions in $M_\D$, $M_\jp$ and $\mrjpd$.
The factor
$N_{\text{rec}}^{(i)} / N_{(\Upsilon(4S)\ E_\gamma < E_\mathrm{cut})}^{(i)}$
is the reconstruction efficiency corrected for the difference between the
full cross section and $\sigma_{\jpdd}^\mathrm{(E_\gamma < E_\mathrm{cut})}$
and for the difference between the number of events for all beam energies
and at the $\Upsilon(4S)$ resonance.
The ratio $\sigma_{\mup \mum}^{(E_\gamma < E_\mathrm{cut})}(\Upsilon(4S)) /
\sigma_{\mup \mum}^\mathrm{(Born)}(\Upsilon(4S))$ is taken from
Ref.~\cite{Kuraev:1985hb}. The ratio
$N_{(\Upsilon(4S)\ E_\gamma < E_\mathrm{cut})}^{(i)} \times
\sigma_{\elp \elm \to \mup \mum}^\mathrm{(Born)}(\Upsilon(4S)) /
\sigma_{\elp \elm \to \mup \mum}^{(E_\gamma < E_\mathrm{cut})}(\Upsilon(4S))$
does not depend on the cutoff energy; it is determined from a fit to its
dependence on $E_\mathrm{cut}$ in the range between 20 and $100\ \mev$.

The background can be subdivided into three classes of events: those with
real $D$ mesons, those with real $\jp$ mesons, and a featureless combinatorial
background.
These components can have a different efficiency dependence on the
MLP output cutoff value $v_0^{(i)}$ as well as different distributions in
$M_\DD$ and angular variables.
To account for these differences, the background region
is divided into three parts with definition indicated in
Fig.~\ref{fig:background_subregions}.
The expected number of the background events is
\begin{equation}
\begin{aligned}
N_\text{bg}^{(i)} = &
\begin{pmatrix}
I_\text{sm}^\text{(SR)} \\
 I_D^\text{(SR)} \\
 I_\jp^\text{(SR)} \\
\end{pmatrix}^{T}
\begin{pmatrix}
I_\text{sm}^\text{(1)} & I_D^\text{(1)} & I_\jp^\text{(1)} \\
I_\text{sm}^\text{(2)} & I_D^\text{(2)} & I_\jp^\text{(2)} \\
I_\text{sm}^\text{(3)} & I_D^\text{(3)} & I_\jp^\text{(3)} \\
\end{pmatrix}^{-1} \\
& \times
\begin{pmatrix}
\epsilon_B^\text{(1)}(v_0^{(i)}) N_\text{(bg rec)}^{(i) (1)} \\
\epsilon_B^\text{(2)}(v_0^{(i)}) N_\text{(bg rec)}^{(i) (2)} \\
\epsilon_B^\text{(3)}(v_0^{(i)}) N_\text{(bg rec)}^{(i) (3)} \\
\end{pmatrix}. \\
\end{aligned}
\label{eq:nbg}
\end{equation}
where each $I$ is a background distribution integral with the superscript
defining the subregion and the subscript identifying the background component,
$N_\text{(bg rec)}^{(i) (j)}$ is the number of reconstructed data
events in the $j$-th background subregion, and
$\epsilon_B^\text{(j)}(v_0^{(i)})$ is the efficiency of the
requirement on the MLP output for the background events
in the $j$-th background subregion.

Four parameters per $D$ channel are determined from the global
selection requirement optimization.
They include the definition of the signal region
($M_\jp^{(i)}$, $M_\D^{(i)}$ and $\mrjpd^{(i)}$) and the MLP output cutoff
value $v_0^{(i)}$.
The MLP output requirement is not used for the two low-background $\D$ decay
channels ($\dcha{1}$ and $\dcha{7}$) because it has no effect.

The resulting signal region definition parameters vary in the range
$33 - 65\ \mevcc$, $8 - 17\ \mevcc$ and $38 - 57\ \mevcc$ for
$M_\jp^{(i)}$, $M_\D^{(i)}$ and $\mrjpd^{(i)}$, respectively. The
finally selected data sample contains $N_\text{obs} = 103$ events.

\begin{figure}
\includegraphics[width=8cm]{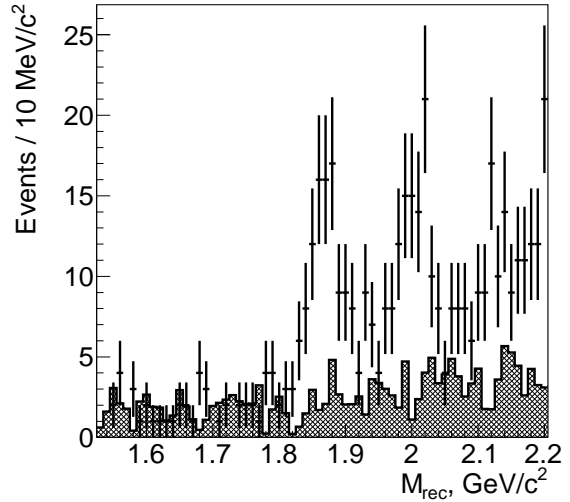}
\caption{The signal and background distributions in $\mrjpd$.
The points with error bars are the data with the channel-dependent signal
region selection requirements defined by Eq.~\eqref{eq:sigreg_jpd} and MLP
output requirements applied.
The hatched histogram is the background.
Note that the definition of the signal region $\mrjpd$ requirement
depends on the $\D$ decay channel.}
\label{fig:mrec}
\end{figure}

After the global optimization, the background fits are performed again;
only events passing the resulting MLP output selection requirement
$v > v_0^{(i)}$ are used.
The weights of the background events for all $D$ channels are determined
as the coefficients of the numbers of background events in the corresponding
background region in Eq.~\eqref{eq:nbg}
(since the MLP output requirement has already been applied,
$\epsilon_B^\text{(j)}(v_0^{(i)}) = 1$). The background event weights are
then used to calculate the background distribution in $\mrjpd$.
The resulting $\mrjpd$ signal and background distributions are
shown in Fig.~\ref{fig:mrec}.

The expected number of background events in the signal region is
$N_\text{bg} = 24.9 \pm 1.1 \pm 1.6$.
The statistical error is calculated from the event weights; it is
found to be 4.5\%.
The systematic error is determined to be 6.3\% and
it includes the error due to the difference of the expected number
of the background events in the training and testing samples
and the statistical error of the background PDF.
The overall error, including both statistical and systematic errors,
is 7.7\%.
The expected fraction of the background events in the signal region
is $b_0 = N_\text{bg} / N_\text{obs} = 0.24 \pm 0.03$.

\section{Amplitude analysis formalism}

We perform an amplitude analysis in a six-dimensional parameter space
\begin{equation}
\Phi = (M_\DD, \thprod, \theta_\jp, \theta_\Xst, \varphi_\lm, \varphi_\D),
\end{equation}
where $\thprod$ is the production angle, $\theta_\jp$ and $\theta_\Xst$ are
the $\jp$ and $\Xst$ helicity angles, respectively, and $\varphi_\lm$ and
$\varphi_\D$ are the $\lm$ and $\D$ azimuthal angles, respectively.
The exact definitions of the angles are given in Appendix~\ref{sec:amplitude}.
In general, the number of dimensions for the process
$\elp \elm \to \lp \lm \D \Db$ is
\begin{equation}
3 N_f - N_c - N_r = 7,
\end{equation}
where $N_f = 4$ is the number of the final state particles,
$N_c = 4$ is the number of conservation relations and
$N_r = 1$ is the number of rotations around the beam axis.
Since the leptons are produced in the decay of the narrow $\jp$ state,
their invariant mass can be ignored, thereby reducing the number of
dimensions by one.

The amplitude of the process $\elp \elm \to \jp \DD$ is represented by a sum
of individual contributions. The default amplitude model
includes a nonresonant amplitude and a
new resonance $\Xst$ in the $\DD$ system with $J^{PC} = 0^{++}$ or $2^{++}$.
Since the $\Xst$ is a resonance in the $\DD$ system, its quantum numbers
are in the ``normal'' series [$P = (-1)^J$]; thus, the asterisk
is added, as for the mesons with open flavor. In the
case of one-photon production, odd values of the $\Xst$ spin
are forbidden by $C$-parity conservation because the $\DD$ pair has
$C=(-1)^{J(\Xst)}$.

The $\Xst$ signal is described by the Breit-Wigner amplitude:
\begin{equation}
A_\Xst(M) = \left(\frac{p(M)}{p(m)}\right)^{L}
\frac{F_L(M)}{m^2 - M^2 - i m \Gamma(M)},
\label{eq:breitwigner}
\end{equation}
where $m$ is the resonance mass, $M$ is the invariant mass of its
daughters, $p(M)$ is the momentum of a daughter particle in the rest frame of
the mother particle calculated at the mother's mass $M$, $L$ is the
decay angular momentum, $F_L(M)$ is the Blatt-Weisskopf form factor
and $\Gamma(M)$ is the mass-dependent width. The Blatt-Weisskopf form factor
is given by~\cite{blattweisskopf}
\begin{equation}
\begin{aligned}
& F_0(M) = 1, \\
& F_2(M) = \sqrt{\frac{z_0^2 + 3 z_0 + 9}{z^9 + 3 z + 9}}, \\
\end{aligned}
\end{equation}
where $z = (p(M) r)^2$ ($r$ being the hadron scale), and $z_0 = (p(m) r)^2$. 
The mass-dependent width is given by
\begin{equation}
\Gamma(M) = \Gamma \left(\frac{p(M)}{p(m)}\right)^{2 L + 1}
\frac{m}{M} F_L^2(M),
\end{equation}
where $\Gamma$ is the resonance width.

The nonresonant amplitude is given by
\begin{equation}
A_\text{NR}(M_\DD) = \sqrt{F_\DD(M_\DD)},
\end{equation}
where $F_\DD(M)$ is the nonresonant-amplitude form factor. Three different
parameterizations are used for $F_\DD(M)$. The default choice is a constant
nonresonant amplitude:
\begin{equation}
F_\DD(M_\DD) = 1.
\end{equation}

Another parameterization is based on Ref.~\cite{Liu:2004ga}. The matrix element
for the process $\elp \elm \to \psi \chi_c$ is integrated over
angle, and $M_\DD$ is used instead of the $\chi_c$ mass.
The result is
\begin{equation}
\begin{aligned}
F_\DD(M_\DD) = \frac{2048}{3 s^7 m_c^2}
\big[&180224 m_c^{10} + 149504 m_c^8 A \\&
+ m_c^6 (44928 A^2 + \frac{1792}{3}B) \\& 
+ m_c^4 (5376 A^3 + \frac{928}{3} A B) \\& 
+ m_c^2 (222 A^4 + 34 A^2 B) \\&
+ (\frac{A^5}{2} - \frac{A^3 B}{6}) \big],
\end{aligned}
\end{equation}
where
\begin{equation}
A = m_\jp^2 + M_\DD^2 - s
\end{equation}
and
\begin{equation}
B = s^2 - 2 s (m_\jp^2 + M_\DD^2) + (m_\jp^2 - M_\DD^2)^2
\end{equation}
with $\sqrt{s} = m_{\Upsilon(4S)}$ and $m_c = 1.5\ \gevcc$.
This nonresonant amplitude model is denoted hereinafter as the
NRQCD model (although, for the actual NRQCD calculation of the
$\elp \elm \to \psi \chi_c$ cross section, the $\chi_c$ mass is set to
$2 m_c$~\cite{Liu:2004ga}).

The third parameterization is based on Ref.~\cite{Bondar:2004sv}.
The matrix element for the process $\jpdd$ is proportional to
$M_\DD^{-4}$ for $4 m_\D^2 \ll M_\DD^2 \ll s$; this dependence is used for
the entire fitting region:
\begin{equation}
F_\DD(M_\DD) = \frac{1}{M_\DD^4}.
\end{equation}
This amplitude model is identified below as the $M_\DD^{-4}$ nonresonant
amplitude.

In the determination of systematic errors,
we also use a model without a nonresonant
contribution but with an additional form factor for the resonant
amplitude~\cite{Hanhart:2012wi}:
\begin{equation}
A_\Xst(M) \to A_\Xst(M) [1 + \beta (M_{\DD} - 2 m_\Dz) ],
\label{eq:alternative_signal}
\end{equation}
where $\beta$ is a form-factor parameter.

The angle-dependent part of the amplitude is calculated using the helicity
formalism (see Appendix~\ref{sec:amplitude}); the result is
\begin{equation}
\begin{aligned}
& A_{\lbeam\,\lambda_\jp\,\lambda_\Xst\,\lambda_\leplep}(\Phi) \\
& \qquad =
  H_{\lambda_\jp\,\lambda_\Xst}
  \dfun{1}{\lbeam}{\lambda_\jp-\lambda_\Xst}{\thprod}
  e^{i \lambda_\jp \varphi_\lm} \\
& \qquad \qquad \times
  \dfun{1}{\lambda_\jp}{\lambda_\leplep}{\theta_\jp}
  e^{i \lambda_\Xst \varphi_\D}
  \dfun{J(\Xst)}{\lambda_\Xst}{0}{\theta_\Xst}, \\
\end{aligned}
\end{equation}
and
\begin{equation}
A_{\lbeam\,\lambda_\leplep}(\Phi) =
\sum\limits_{\substack{\lambda_\jp = -1, 0, 1 \\ \lambda_X = -J(X), ..., J(X)}}
A_{\lbeam\,\lambda_\jp\,\lambda_\Xst\,\lambda_\leplep}(\Phi)
\end{equation}
where $H_{\lambda_\jp\,\lambda_\Xst}$ is the production helicity
amplitude, $\dfun{j}{m_1}{m_2}{\theta}$ are Wigner $d$-functions,
$\lambda$ is the helicity of the particle specified by the index,
$\lbeam$ is the sum of helicities of the beam electron and positron and
$\lambda_\leplep$ is the sum of helicities of the $\jp$ decay products.

The signal density function is
\begin{equation}
S(\Phi) = \sum\limits_{\substack{\lbeam = -1, 1\\\lambda_\leplep = -1, 1}} 
\Big| \sum\limits_\Xst A_{\lbeam\,\lambda_\leplep}(\Phi) A_\Xst(M_\DD)
\Big|^2,
\end{equation}
where $\Xst$ is any contribution to the signal ($\Xst$ resonance or
nonresonant amplitude). The resolution in $M_\DD$ may be taken into account
by substituting
\begin{equation}
\begin{aligned}
S(\Phi) \to \sum_{i_g = -n_g}^{n_g} & R_{M_\DD}(M_\DD, i_g \Delta M_\DD)
\Delta M_\DD \\
& \times S(M_\DD + i_g \Delta M_\DD, ...), \\
\end{aligned}
\end{equation}
where $R_{M_\DD}$ is the resolution in $M_\DD$, $\Delta M_\DD$ is the distance
between grid points and $i_g$ is the grid point index.
The resolution in $M_\DD$
is ${\sim}12\ \mevcc$ for $M_\DD$ range between 5 and 6 $\gevcc$;
it is smaller
for lower $\DD$ masses. The resolution is much smaller than the width of the
$M_\DD$ structure (see Fig.~\ref{fig:sigfit}) and can be ignored for the
default fit; nevertheless, it is taken into account since alternative
models with relatively narrow states $X(3915)$ or $\chi_{c2}(2P)$ are
considered.

The construction of the likelihood function follows Ref.~\cite{Garmash:2004wa}.
The function to be minimized is
\begin{equation}
F = - 2 \sum\limits_{k}
\ln\Big( (1-b)\frac{S(\Phi_k)}
{\sum\limits_{l}S(\Phi_l)}
 +b\frac{B(\Phi_k)}{\sum\limits_{l}B(\Phi_l)}\Big) +
\frac{(b - b_0)^2}{\sigma_{b_0}^2},
\label{eq:fcn}
\end{equation}
where $b$ is the fraction of the background events, $B(\Phi)$
is the background density in the signal region
determined from the fit to the background, $b_0$ is the expected
background fraction and $\sigma_{b_0}^2$ is its error.
The index $k$ runs over data events; the index $l$
runs over MC events generated uniformly over the phase space
and reconstructed using the same selection requirements as in data.
This procedure takes into account the nonuniformity of the
reconstruction efficiency but requires
a parameterization of the background shape.
For the background, the likelihood is
\begin{equation}
F = - 2 \sum\limits_{k}
w_k \ln\Big(\frac{B(\Phi_k)}{\sum\limits_{l}B(\Phi_l)}\Big),
\label{eq:bg_fcn}
\end{equation}
where $w_k$ is the weight of the $k$-th background event,
calculated as described in Sec.~\ref{sec:selection_optimization}.

\section{Fit to the data}

\subsection{Fit to the background}

\begin{figure*}
\includegraphics[width=5.8cm]{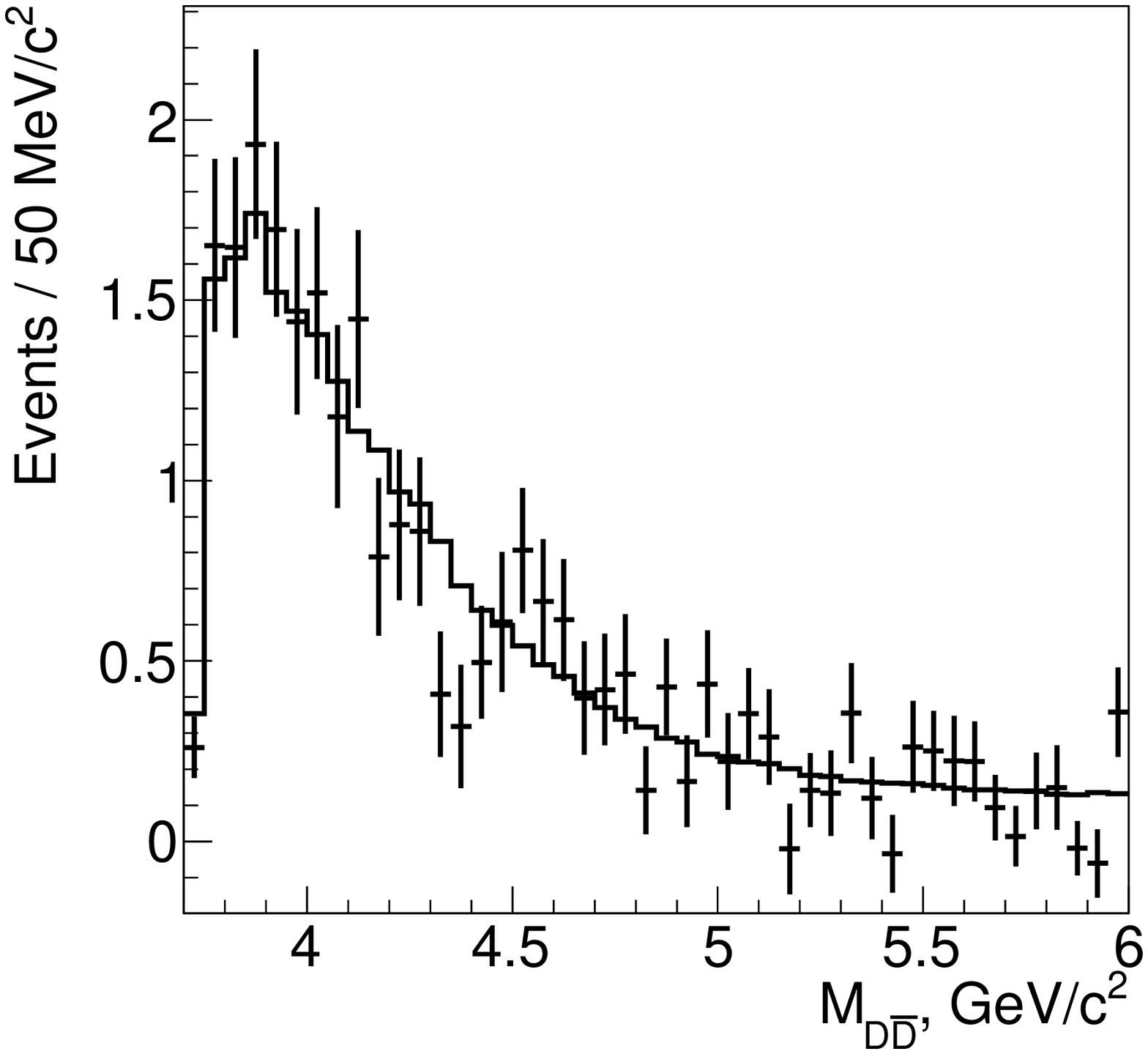}
\includegraphics[width=5.8cm]{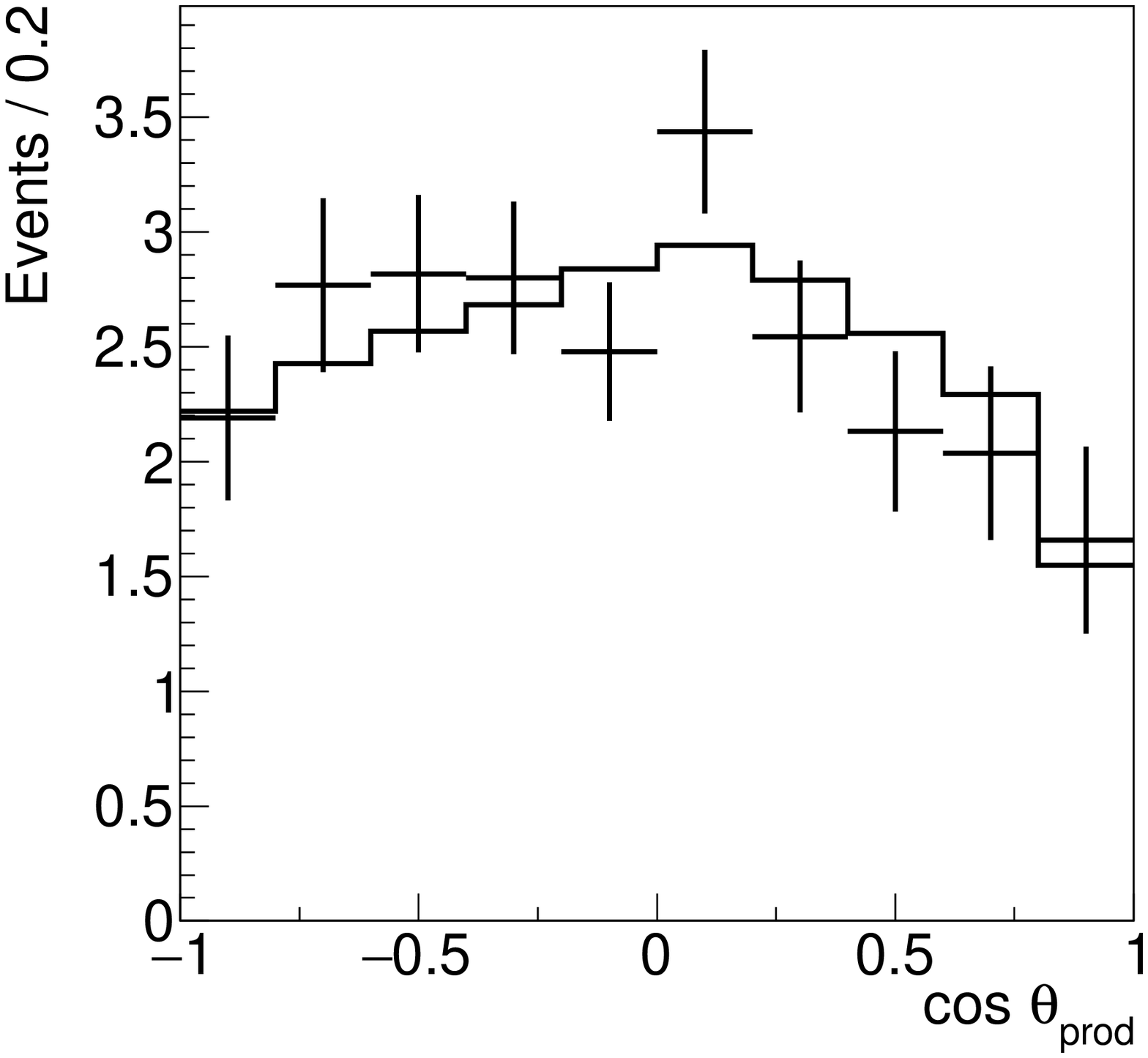}
\includegraphics[width=5.8cm]{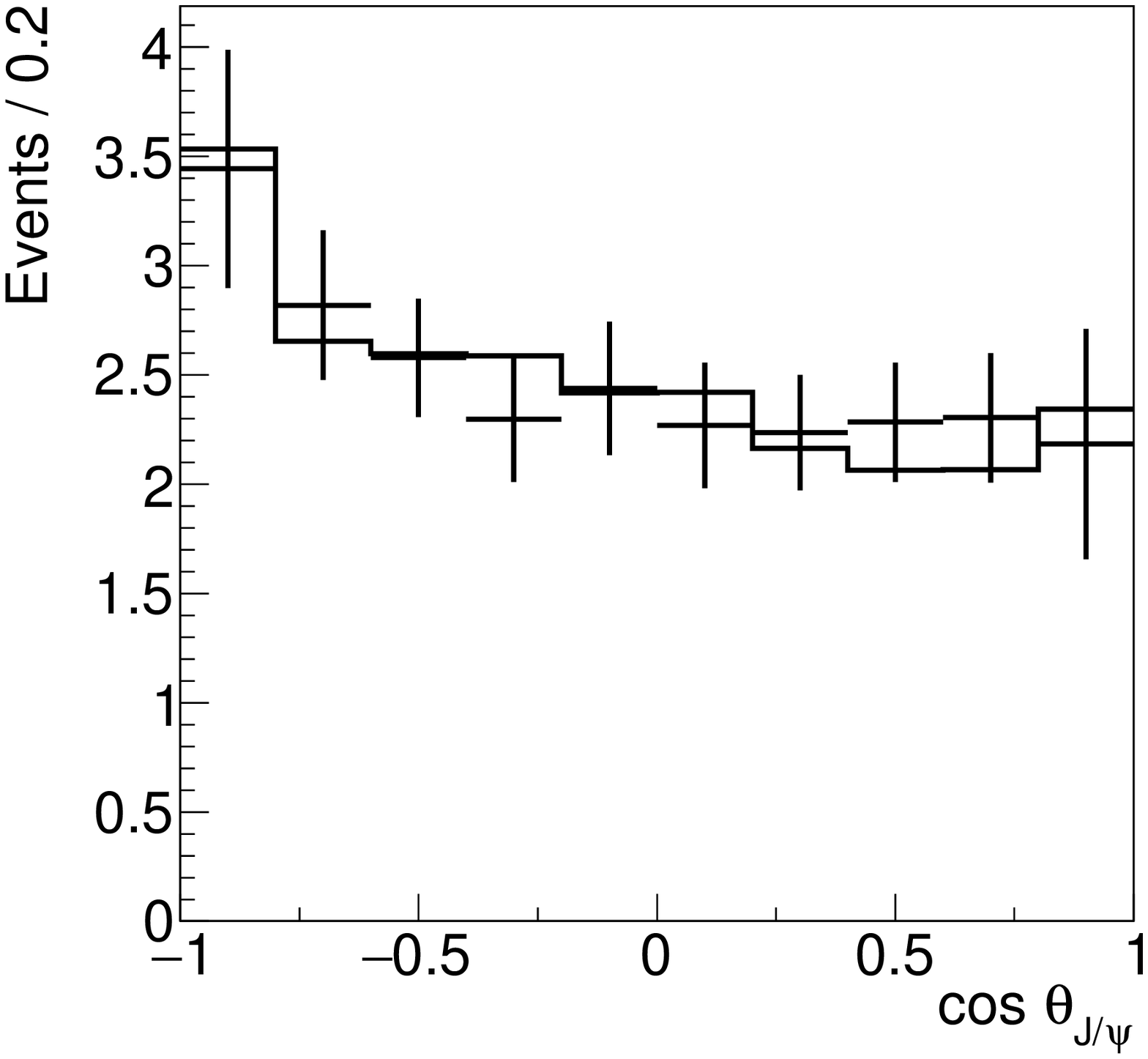}
\includegraphics[width=5.8cm]{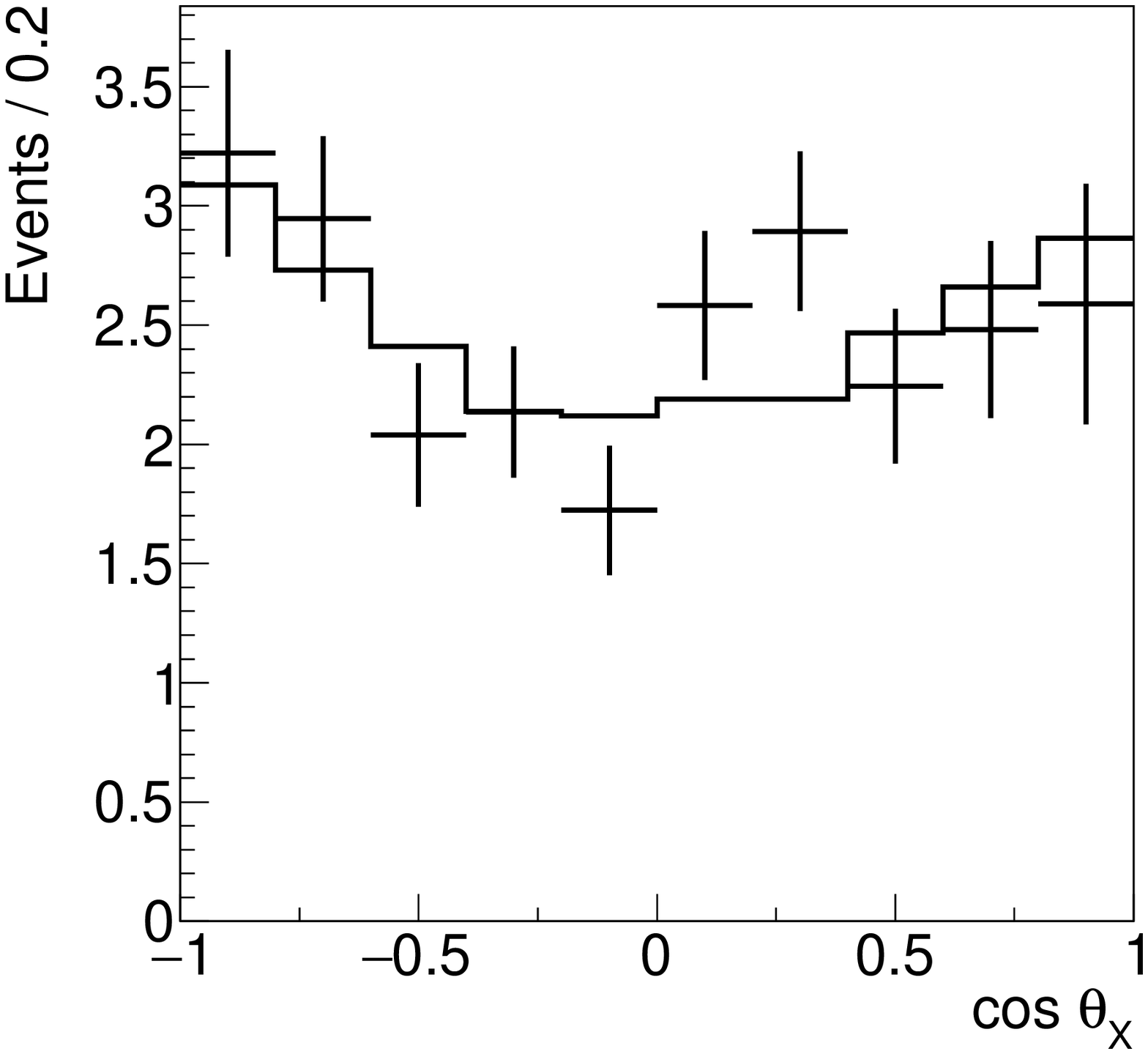}
\includegraphics[width=5.8cm]{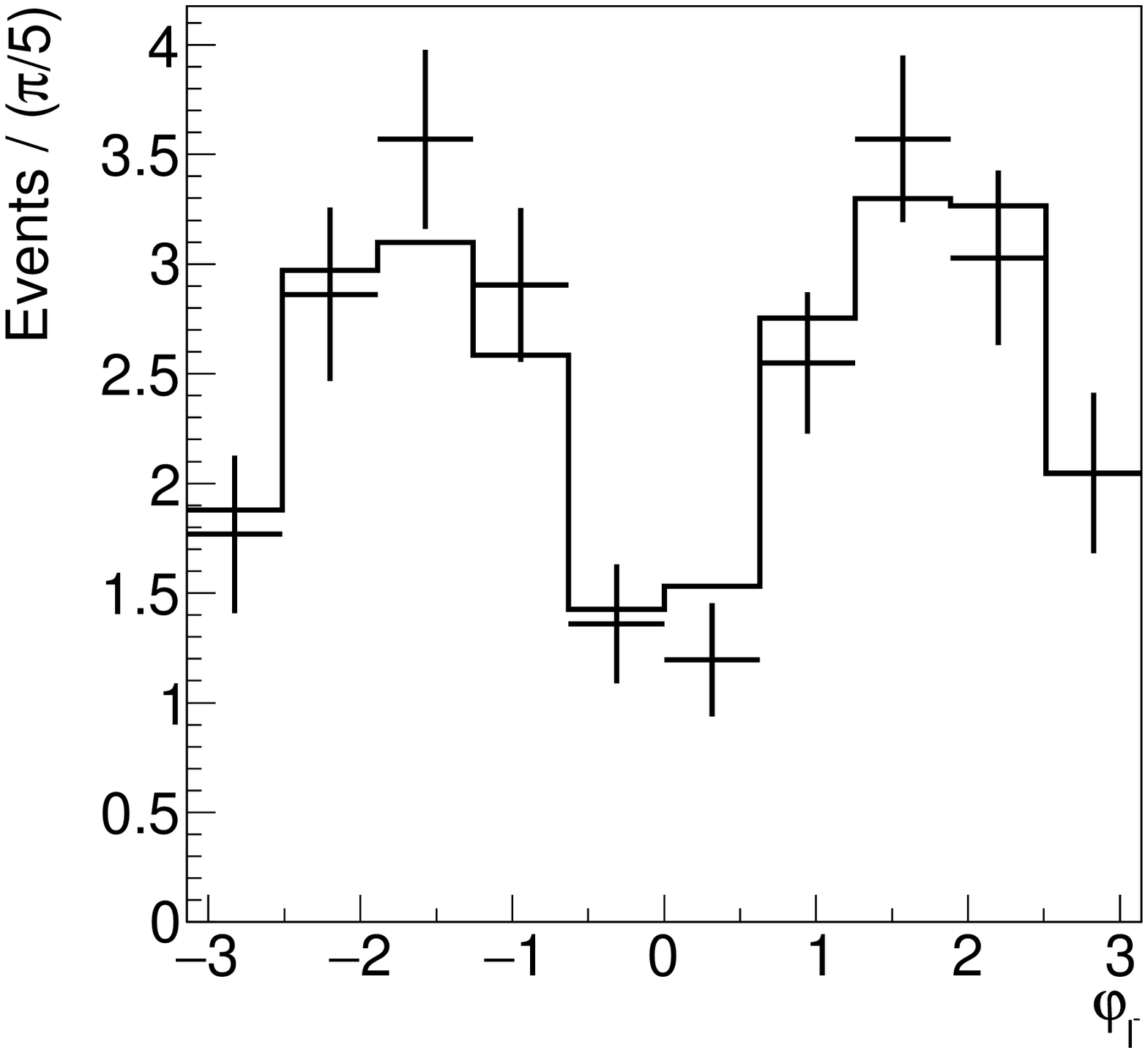}
\includegraphics[width=5.8cm]{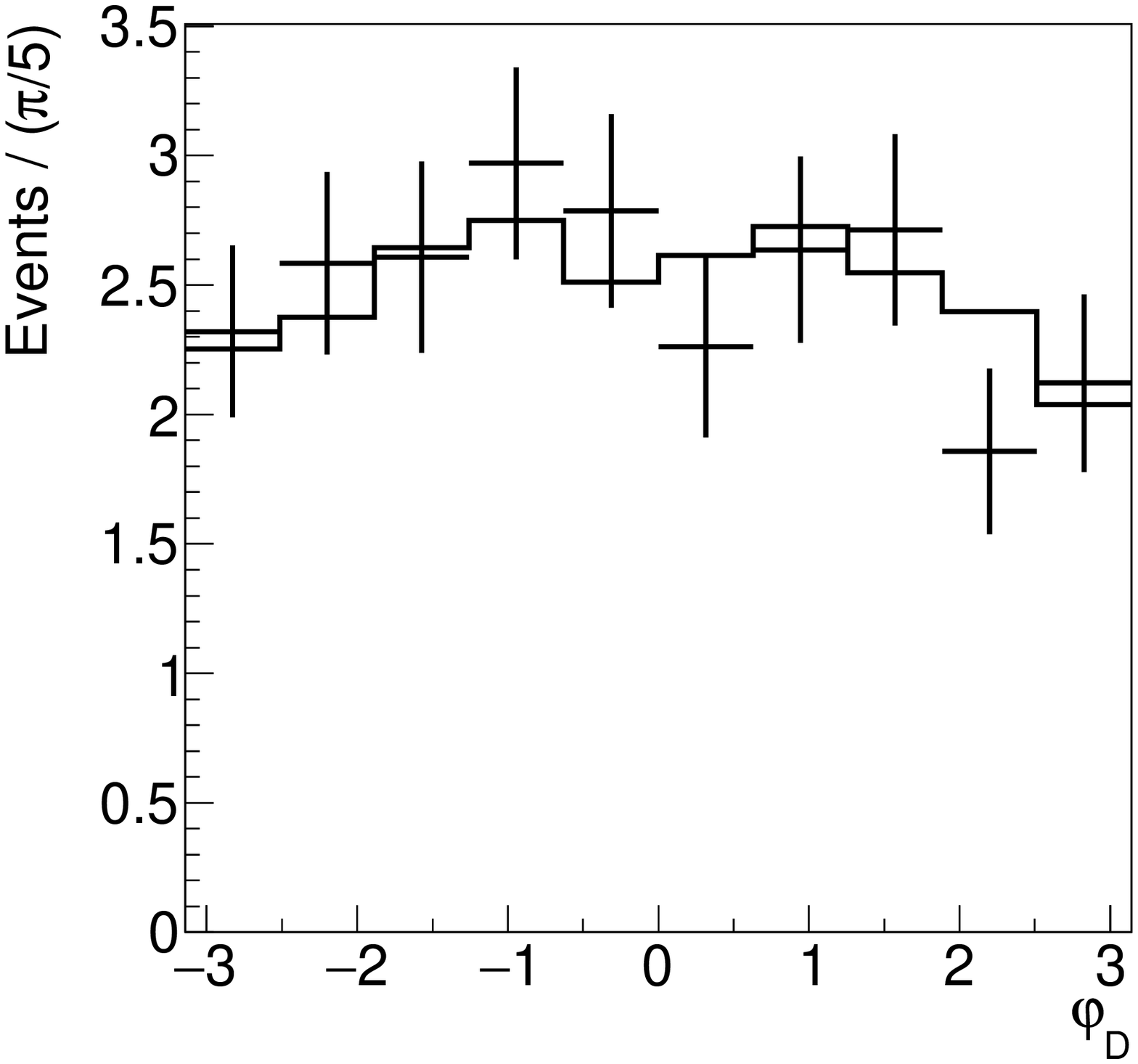}
\caption{Projections of the background fit results onto $M_\DD$ and
angular variables. The points with error bars are
data and the solid line is the fit result.}
\label{fig:bgfit}
\end{figure*}

The background density function is given by
\begin{equation}
B(\Phi) = \frac{P_2(\Phi) \left(\exp(-\xi [p(M_\DD)]^2) + \eta\right)
[p(M_\DD)]^{-\zeta}}
{[1 - u_1 (\cos^2 \thprod)^{v_1}][1 - u_2 (\cos^2 \theta_\jp)^{v_2}]},
\end{equation}
where $P_2(\Phi)$ is a six-dimensional second-order polynomial and
$u_1$, $u_2$, $v_1$, $v_2$, $\xi$, $\eta$ and $\zeta$ are coefficients.
The projections of the fit results onto the individual variables are
shown in Fig.~\ref{fig:bgfit}.

\subsection{Fit to the signal}

\begin{figure*}
\includegraphics[width=5.8cm]{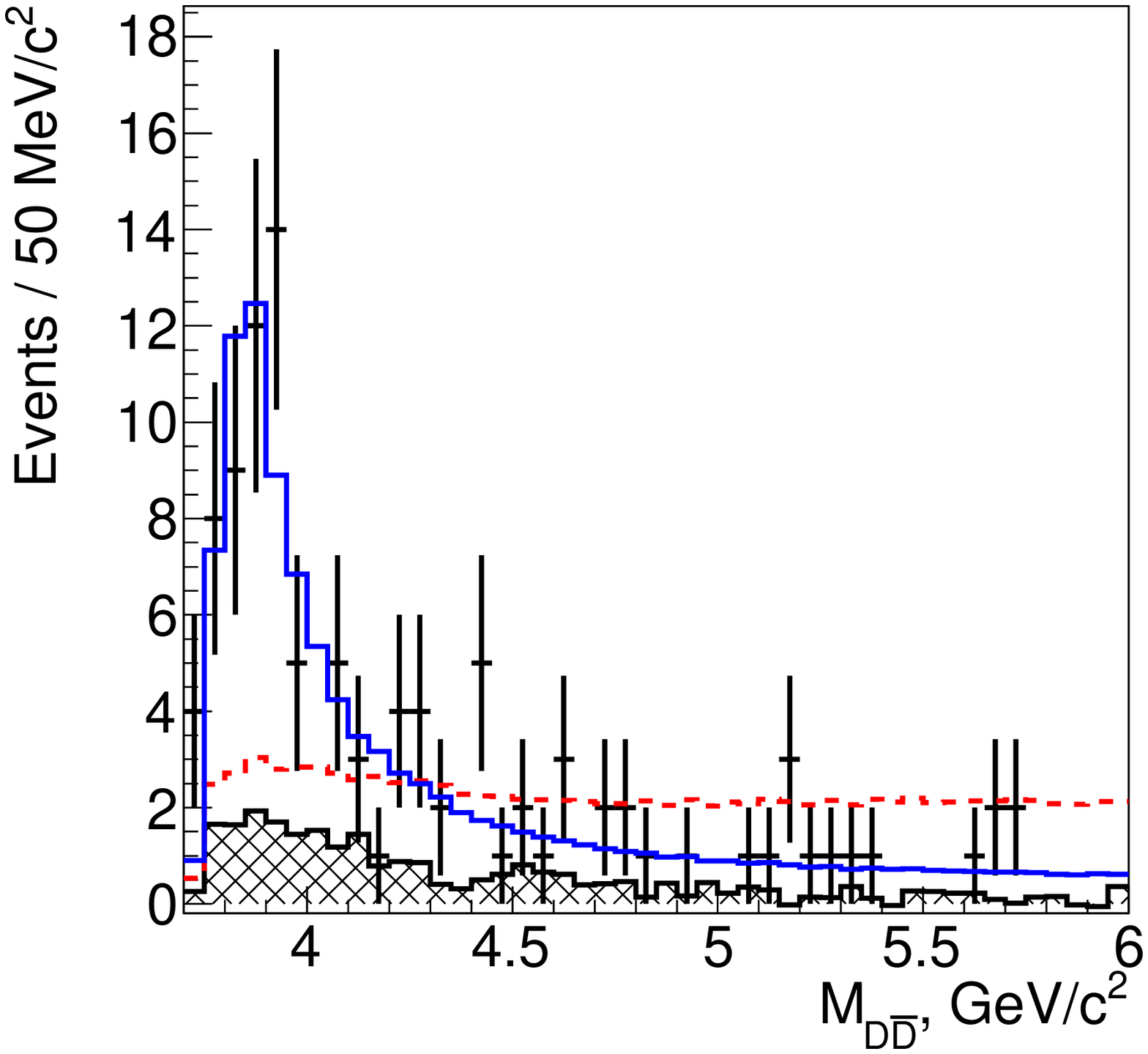}
\includegraphics[width=5.8cm]{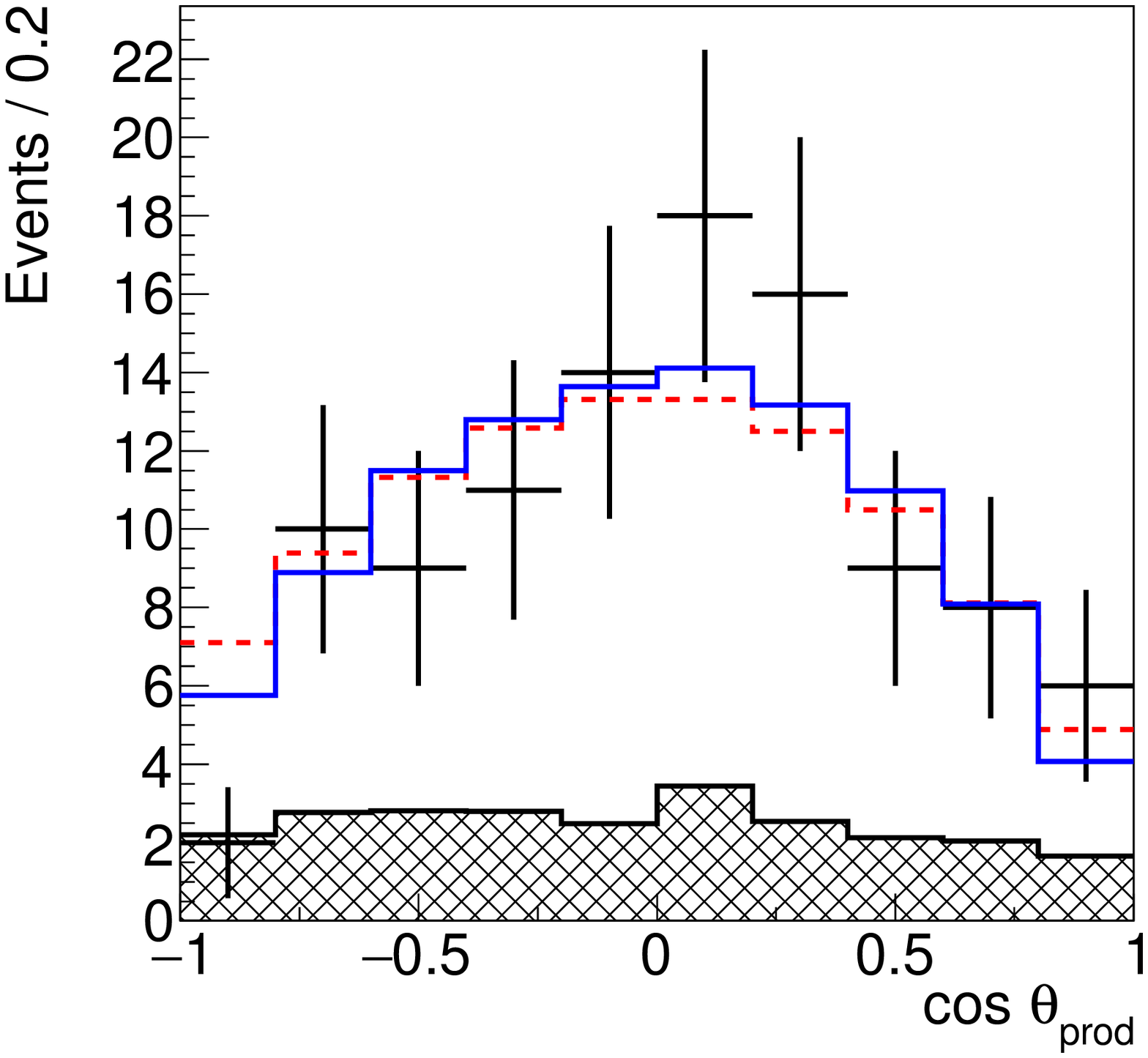}
\includegraphics[width=5.8cm]{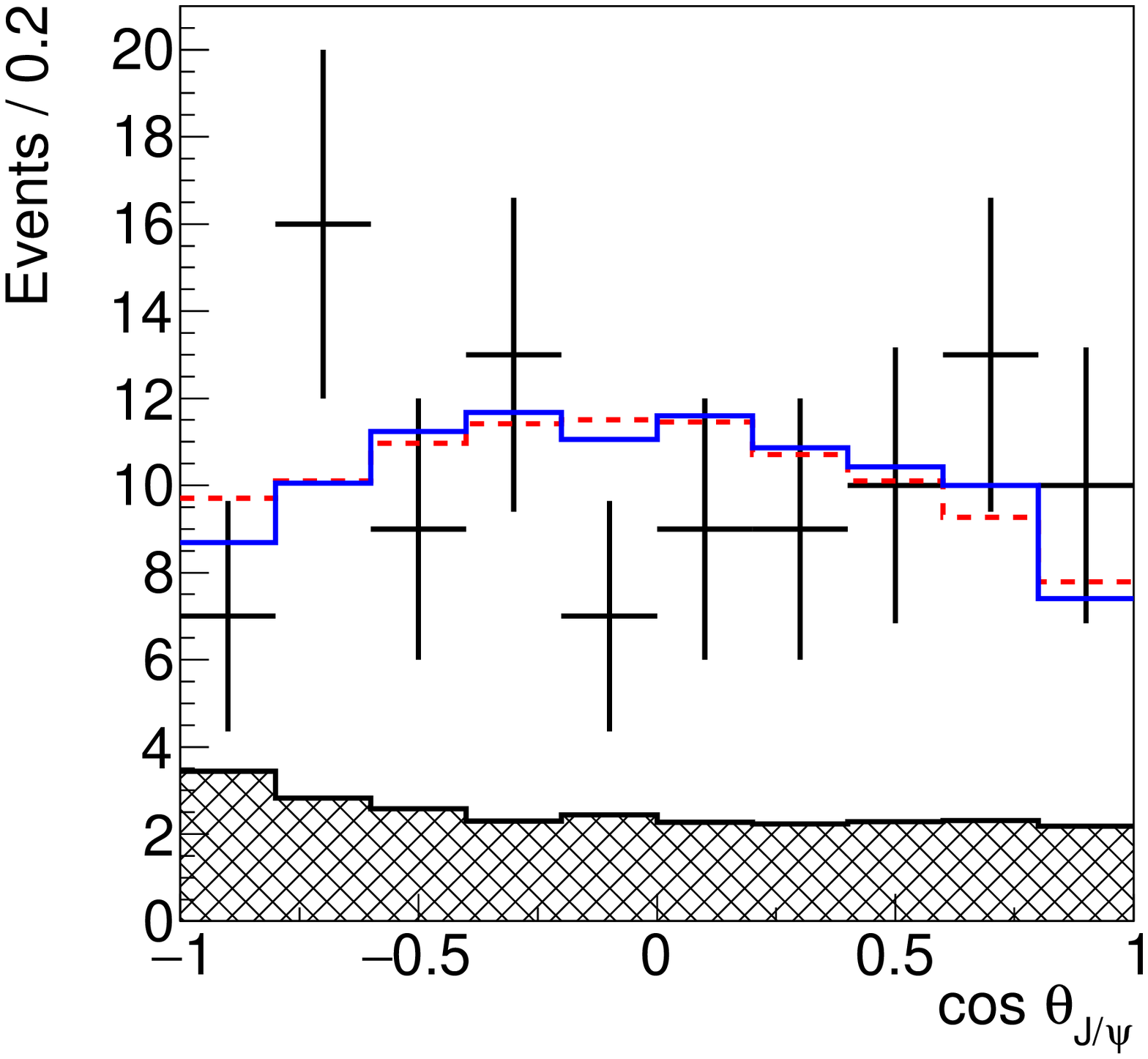}
\includegraphics[width=5.8cm]{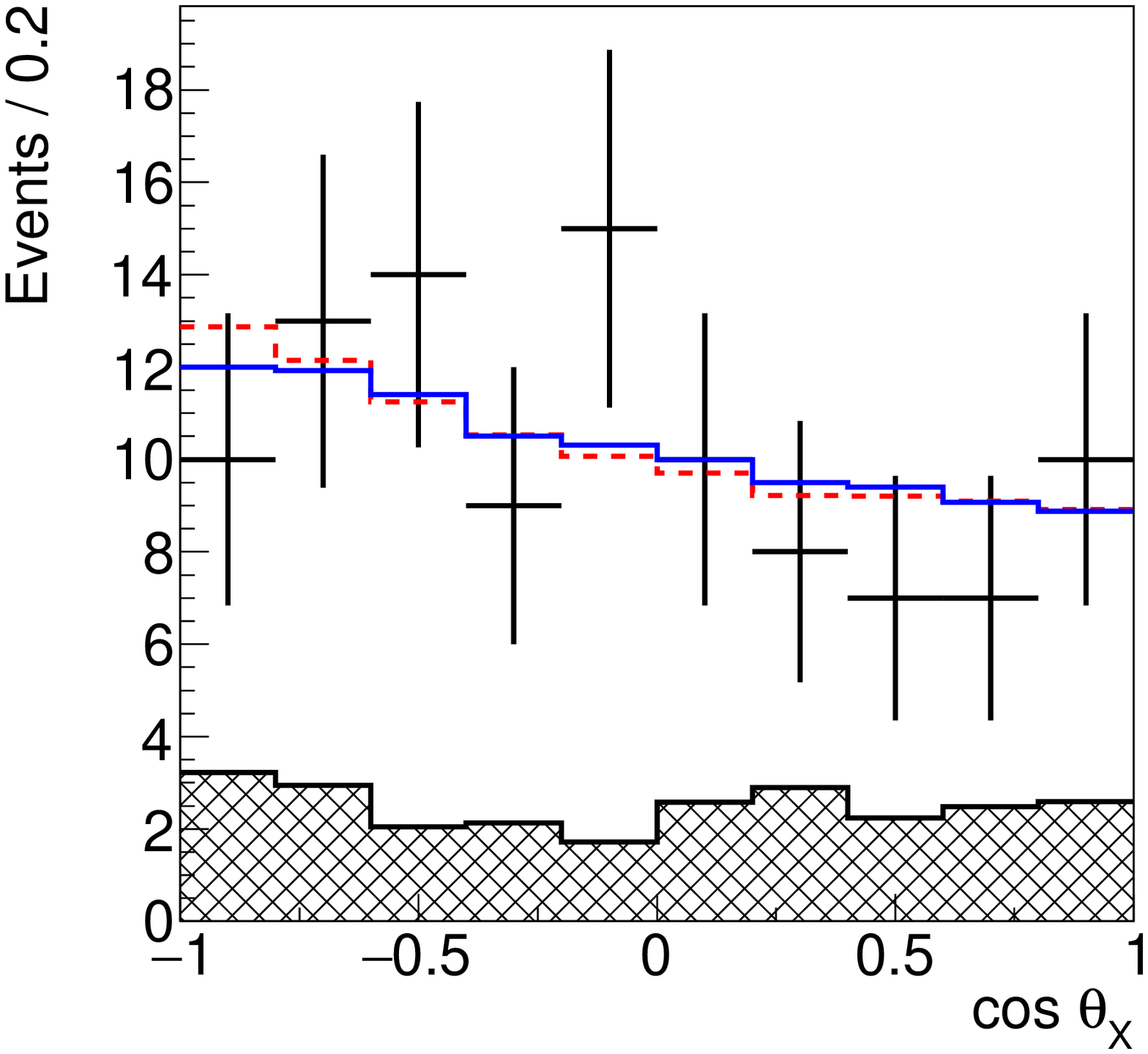}
\includegraphics[width=5.8cm]{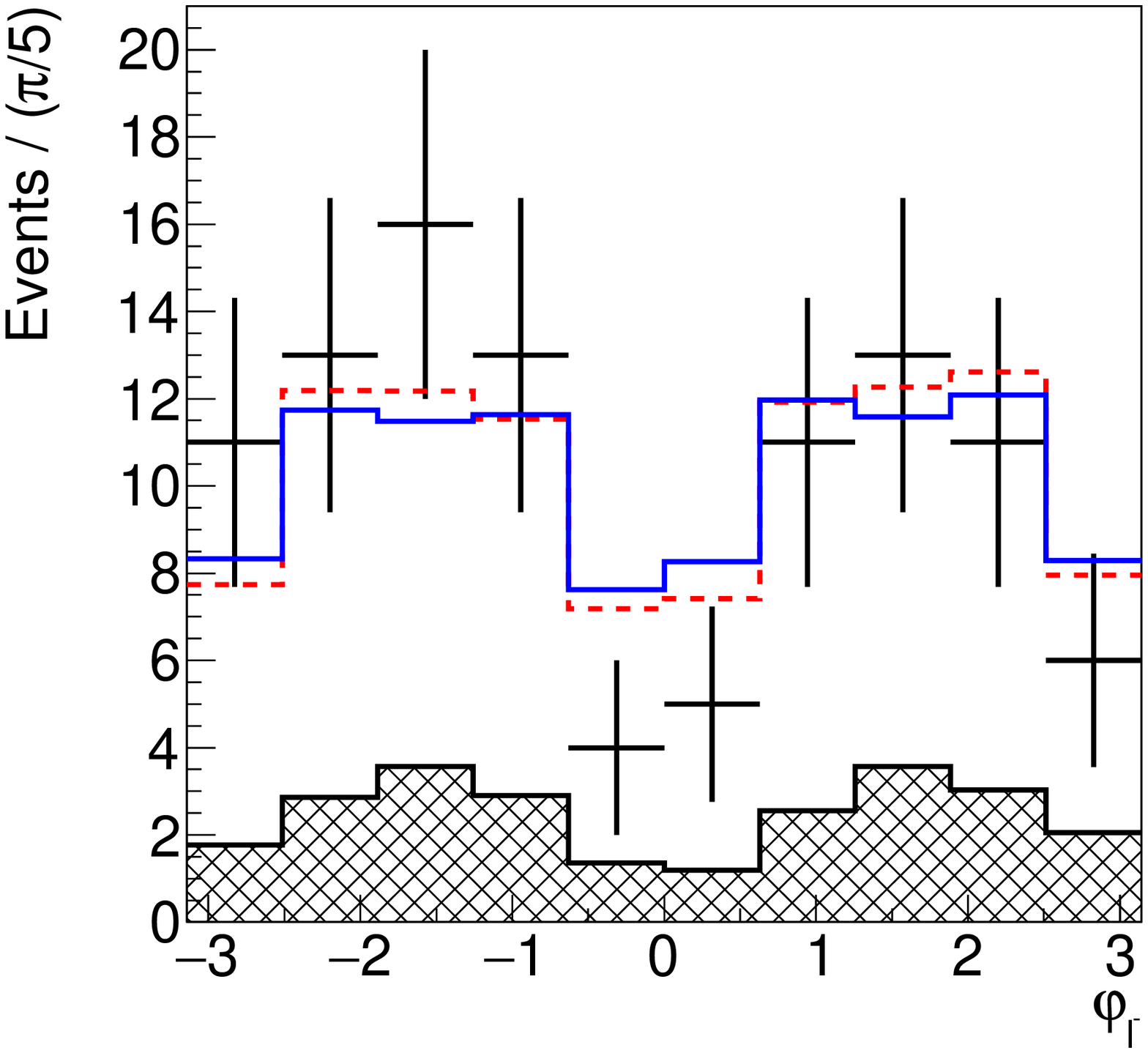}
\includegraphics[width=5.8cm]{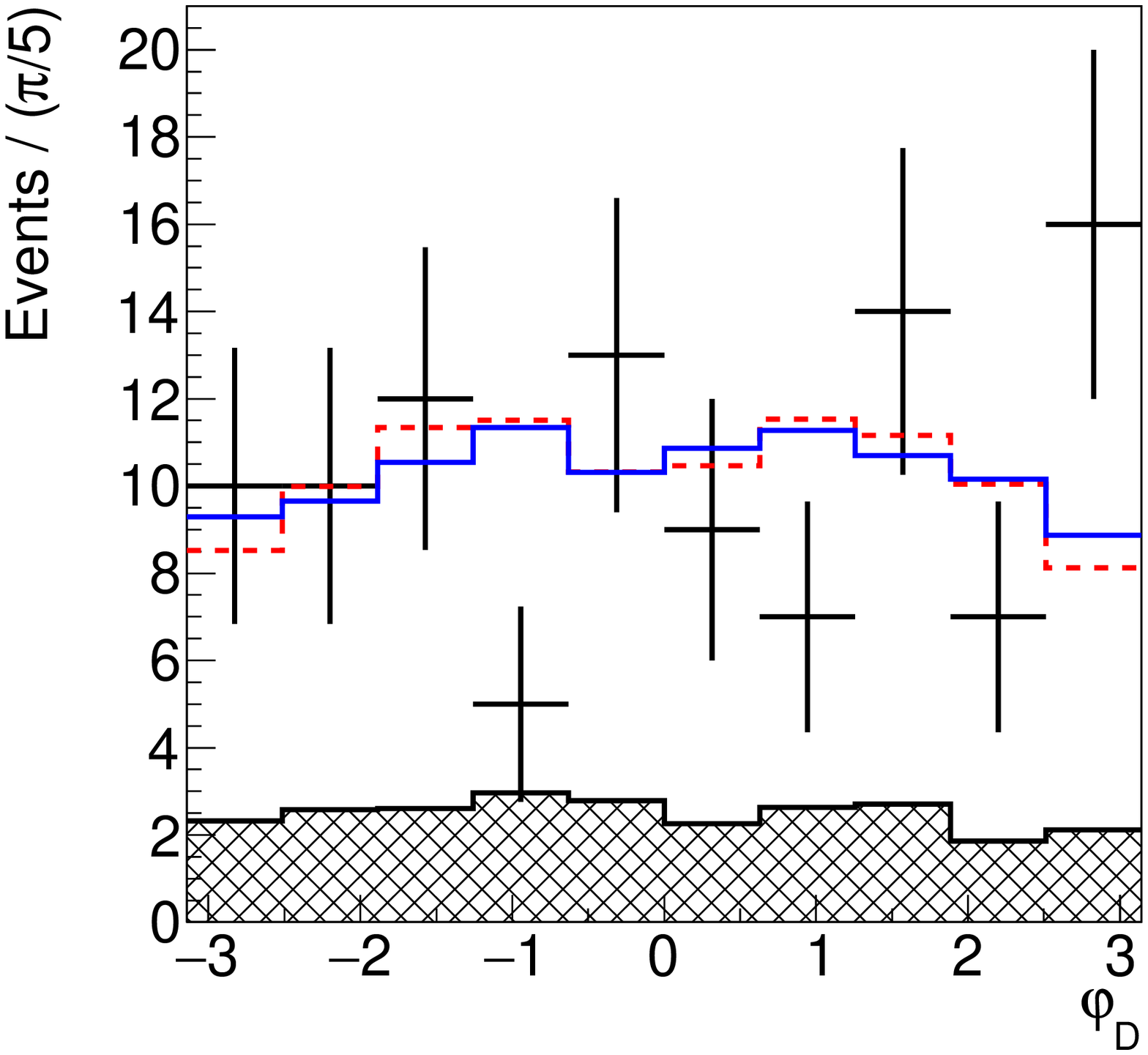}
\caption{Projections of the signal fit results in the default model
onto $M_\DD$ and
angular variables. The points with error bars are the data, the hatched
histograms are the background, the blue solid
line is the fit with a new $\Xst$ resonance ($J^{PC}=0^{++}$) and
the red dashed line is the fit with nonresonant amplitude only.}
\label{fig:sigfit}
\end{figure*}

The result of the fit to the signal events in the default model are listed in
Table~\ref{tab:fitres} for two $\Xst$ quantum-number hypotheses:
$J^{PC}=0^{++}$ and $J^{PC}=2^{++}$. There are three solutions for
the $2^{++}$ hypothesis; all solutions have very similar likelihood values
($|\dlnl| < 1$).
A significant resonance is observed for both the $0^{++}$ and $2^{++}$
hypotheses;
the $0^{++}$ hypothesis is preferred.
The significance is calculated
using Wilks' theorem~\cite{Wilks:1938dza} with six and twelve degrees of
freedom for the $0^{++}$ and $2^{++}$ hypotheses, respectively.
The global significance calculated using the method described in
Ref.~\cite{Aaij:2014jqa} is $8.5\sigma$ for the $0^{++}$ hypothesis.
Thus, we observe
a new charmoniumlike state that is referred to hereinafter
as the $\xst{3860}$.
The projections of the fit results
onto $M_{\DD}$ and the angular variables for the case of the $\xst{3860}$
with $J^{PC}=0^{++}$ are shown in Fig.~\ref{fig:sigfit} and the amplitudes
are shown in Table~\ref{tab:sigfit}. The helicity amplitudes
$H_{1\,0}$ and $H_{0\,0}$ almost exactly coincide (with large error). This is
consistent with pure $S$-wave production of the $\xst{3860}$ [see
Eq.~\eqref{eq:helpwa}].

\begin{table}
\caption{Fit results in the default model (constant nonresonant amplitude).
For the $2^{++}$ hypothesis, there are three solutions.}
\begin{tabular}{c|c|c|c}
\hline\hline
$J^{PC}$ & Mass, $\mevcc$ & Width, $\mev$ & Significance \\
\hline
$0^{++}$ & $3862\err{26}{32}$ & $201\err{154}{67}$ & $9.1\sigma$ \\
$2^{++}$ & $3879\err{20}{17}$ & $171\err{129}{62}$ & $8.0\sigma$ \\
$2^{++}$ & $3879\err{17}{17}$ & $148\err{108}{50}$ & $8.0\sigma$ \\
$2^{++}$ & $3883\err{26}{24}$ & $227\err{201}{125}$ & $8.0\sigma$ \\
\hline\hline
\end{tabular}
\label{tab:fitres}
\end{table}

\begin{table}
\caption{The $\xst{3860}$ ($J^{PC}=0^{++}$) amplitudes in the default model.}
\begin{tabular}{c|c}
\hline\hline
Amplitude & Value \\
\hline
$\Re H_{0\,0}$ & 1 (fixed) \\
$\Im H_{0\,0}$ & 0 (fixed) \\
$\Re H_{1\,0}$ & $1.00 \pm 0.38$ \\
$\Im H_{1\,0}$ & $0.01 \pm 0.93$ \\
\hline\hline
\end{tabular}
\label{tab:sigfit}
\end{table}

Other known states with $J^{PC} = 0^{++}$ or $2^{++}$ may also contribute to
this process. We consider the states with mass ${\sim}3.9\ \gevcc$
[$X(3915)$ and $\chi_{c2}(2P)$] and the states observed in double-charmonium
production [$X(3940)$ and $X(4160)$]. Note that the $X(3940)$ decays to
$D^* \bar{D}$~\cite{Abe:2007jna,Abe:2007sya} and it consequently may be
observed in the process $\jpdd$ only if its quantum numbers are
$J^{PC} = 2^{++}$.
As in Ref.~\cite{Abe:2007sya}, because of low statistics we do not consider
the possibility that the $\xst{3860}$ peak is due to more than one state.
We check whether the $\xst{3860}$ is compatible with the states listed above
by performing a fit with Gaussian constraints
to the known resonance parameters~\cite{Agashe:2014kda}
on the $\Xst$ mass and width.
For known states with $J^{PC}=2^{++}$, the exclusion levels are calculated
from MC pseudoexperiments similarly
to the comparison of $J^{PC}=0^{++}$ and $2^{++}$ hypotheses described
in Sec.~\ref{sec:jpc_comparison};
for $0^{++}$ states, the exclusion levels are calculated as $\sqrt{\dlnl}$.
The calculation is performed for all nonresonant amplitude
models. The results are listed in Table~\ref{tab:known_exclusion}. The
$X(3915)$ is excluded at $3.3\sigma$ and $4.9\sigma$ in the case of
$J^{PC}=0^{++}$ and $2^{++}$, respectively. Other known states are
excluded at more than $5\sigma$.
In addition, we compare the $\xst{3860}$
and the lattice prediction for the $\chicp$ provided
in Ref.~\cite{Lang:2015sba}.
The parameters $M = 3966 \pm 20\ \mevcc$ and $\Gamma = 67 \pm 18\ \mev$
are used. The comparison is performed in the same way as those for the
experimentally known states with $J^{PC}=0^{++}$; the results are
shown in Table~\ref{tab:known_exclusion}. The difference of the
$\xst{3860}$ and predicted $\chicp$ parameters is at $2.7\sigma$ level.
Note that the systematic errors have not been determined in
Ref.~\cite{Lang:2015sba}; thus, the actual level of disagreement
should be less than $2.7\sigma$.

\begin{table}
\caption{Comparison of the $\xst{3860}$ and known charmoniumlike states:
exclusion levels for the hypotheses of the $\xst{3860}$ being the indicated
state for different nonresonant amplitude models.}
\begin{tabular}{c|c|c|c|c}
\hline\hline
\multirow{2}{*}{State} & \multirow{2}{*}{$J^{PC}$} &
\multicolumn{3}{|c}{Nonresonant amplitude} \\
\cline{3-5}
& & Constant & NRQCD & $M_{DD}^{-4}$ \\
\hline
$X(3915)$ & $0^{++}$ & $5.2\sigma$ & $4.3\sigma$ & $3.3\sigma$ \\
$X(3915)$ & $2^{++}$ & $6.1\sigma$ & $6.1\sigma$ & $4.9\sigma$ \\
$\chi_{c2}(2P)$ & $2^{++}$ & $6.8\sigma$ & $7.0\sigma$ & $6.2\sigma$ \\
$X(3940)$ & $2^{++}$ & $6.0\sigma$ & $5.6\sigma$ & $5.2\sigma$ \\
$X(4160)$ & $0^{++}$ & $6.8\sigma$ & $6.3\sigma$ & $5.8\sigma$ \\
$X(4160)$ & $2^{++}$ & $10.7\sigma$ & $11.0\sigma$ & $13.5\sigma$ \\
$\chicp$ (lattice) & $0^{++}$ & $4.3\sigma$ & $3.6\sigma$ & $2.7\sigma$ \\
\hline\hline
\end{tabular}
\label{tab:known_exclusion}
\end{table}

\subsection{Systematic uncertainties}

The systematic uncertainties of the $\xst{3860}$ mass and width
are listed in Table~\ref{tab:syserrmasswidth}.
One source of mass and width systematic error is
the systematic uncertainty due to the nonresonant amplitude model dependence.
We perform a fit with all nonresonant amplitude models and consider the
maximal variations of the $\xst{3860}$ mass and width as the systematic
uncertainty. Note that there are two solutions for both fits with
the NRQCD and $M_{\DD}^{-4}$ nonresonant amplitudes in the case of
$J^{PC} = 0^{++}$; all solutions are included into the calculation of the
maximal variations. The systematic error related to the alternative signal
model defined in Eq.~\eqref{eq:alternative_signal}
is included separately.

\begin{table}
\caption{Systematic errors for the $\xst{3860}$ mass (in $\mevcc$) and width
(in $\mev$).}
\begin{tabular}{c|c|c}
\hline\hline
Error source & Mass & Width \\
\hline
Nonresonant amplitude model & $\err{40.2}{0.0}$ & $\err{0.0}{82.0}$ \\
Signal model & $\err{0.0}{10.2}$ & $\err{0.0}{4.0}$ \\
Fit bias & --- & $\err{32.6}{0}$ \\
Optimization & $\err{0.0}{3.1}$ & $\err{71.1}{0.0}$ \\
Background mass calculation & $\err{0.0}{7.9}$ & $\err{40.0}{0.0}$ \\
$\D$ mass & $\pm 0.2$ & --- \\
\hline
Total & $\err{40.2}{13.3}$ & $\err{87.9}{82.1}$ \\
\hline\hline
\end{tabular}
\label{tab:syserrmasswidth}
\end{table}

A different fit-related systematic uncertainty source
is the fit bias. The fit bias is estimated from the mass and width
distributions in MC pseudoexperiments generated in accordance with the default
fit result. We find that the position of the peak of the mass distribution
is in good agreement with the fit result in data, while the position of
the peak of the width distribution is shifted from the value in data. Thus,
a fit bias uncertainty is assigned only to the $\xst{3860}$ width.

Another systematic uncertainty source considered in the
previous analysis~\cite{Abe:2007sya} is the variation of selection requirements.
There is no straightforward analog of this uncertainty in the new analysis
because of the complex selection optimization procedure. However, its last
stage (the global optimization) can be varied. We change the target
significance $a$ in Eq.~\eqref{eq:optfun} from its default value (5) by $\pm 1$
unit,
repeat the global optimization and fitting with the new target significance
and treat the variations of the $\xst{3860}$ mass and width as the systematic
uncertainty related to the optimization.

The default $M_\DD$ calculation procedure for the background
events is the same as for the events in the signal region. We perform a
mass-constrained fit to the $\jp$ and $\D$ candidates with the candidate mass
fixed at $m_\jp$ and $m_\D$, respectively. Alternatively, the background region
is divided into smaller rectangular $(M_\jp,M_\D)$ subregions.
The candidate mass is fixed at the center of its subregion. After the fit
with $\mrjpd$ constrained to $m_\D$, the resulting value of $m_\DD$
is shifted in such a way that the threshold of the new distribution is at
$2 m_\D$. The background distribution is then calculated in the same way as for
the default mass calculation method; the difference in the fit results is
considered as the background mass calculation method systematic uncertainty.

The mass also has a systematic error due to the uncertainty in the mass
of the $\xst{3860}$ decay products (the $D$ mesons).

The $\xst{3860}$ global significance for alternative models is shown in
Table~\ref{tab:systsign}. The minimal global significance of the $\xst{3860}$
is found to be $6.5\sigma$. Thus, the new state $\xst{3860}$ is clearly
observed, accounting for both statistical and systematic uncertainties.

\begin{table}
\caption{The $\xst{3860}$ global significance for alternative models.}
\begin{tabular}{c|c}
\hline\hline
Model & Significance \\
\hline
Default (constant nonresonant) & $8.5\sigma$ \\
NRQCD nonresonant              & $7.6\sigma$ \\
$M_{\DD}^{-4}$ nonresonant     & $6.5\sigma$ \\
Background mass calculation    & $8.4\sigma$ \\
Optimization ($a=4$)           & $8.1\sigma$ \\
Optimization ($a=6$)           & $8.1\sigma$ \\
\hline\hline
\end{tabular}
\label{tab:systsign}
\end{table}

\subsection{Comparison of $J^{PC}$ hypotheses}
\label{sec:jpc_comparison}

The $\xst{3860}$ quantum number hypotheses $J^{PC}=0^{++}$ and $2^{++}$ are
compared using MC simulation. We generate MC pseudoexperiments in
accordance with the fit result with the $2^{++}$ $\xst{3860}$ signal in data
and fit them with the $2^{++}$ and $0^{++}$ signals.
The resulting distribution of
$\dlnl = (-2\ln L)_{J^{PC}=2^{++}} - (-2\ln L)_{J^{PC}=0^{++}}$
is fitted to an asymmetric Gaussian function and the $p$-value
is calculated as the integral of the fitting function normalized to 1
from the value of $\dlnl$ in data to $+\infty$.
If there are multiple solutions for the $2^{++}$ hypothesis in data (this is
the case for the default model), then the exclusion level depends on the
solution. The exclusion level calculation is performed for all solutions.
The comparison the $\xst{3860}$ quantum number hypotheses
includes an additional model, where the Blatt-Weisskopf form factor hadron
scale $r$ is allowed to vary between $0$ and $10\ \gev^{-1}$ (this model is not
included in the calculation of the systematic errors of the $\xst{3860}$
mass and width because, for the preferred hypothesis $J^{PC}=0^{++}$, the
Blatt-Weisskopf form factor is $F_0 = 1$).
The results are presented in Table~\ref{tab:excl}.
The $J^{PC}=0^{++}$ hypothesis is favored over the $2^{++}$
hypothesis at the level of $2.5\sigma$.

\begin{figure}
\includegraphics[width=8cm]{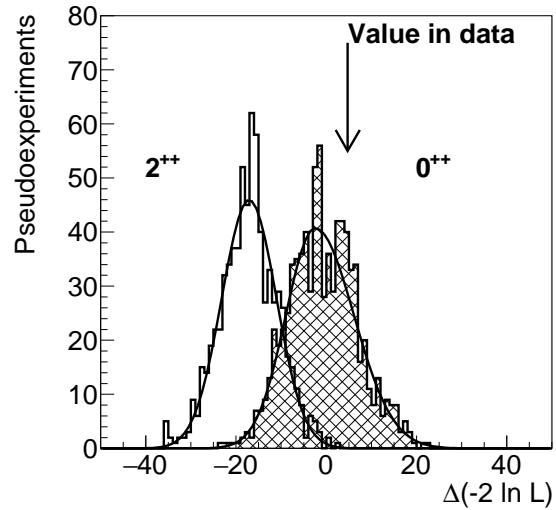}
\caption{Comparison of the $0^{++}$ and $2^{++}$ hypotheses in the default
model (constant nonresonant amplitude).
The histograms are distributions of $\dlnl$ in MC pseudoexperiments
generated in accordance with the fit results with $2^{++}$
(open histogram) and $0^{++}$ (hatched histogram) signals.
The $2^{++}$ pseudoexperiments are generated for the second solution (see
Table~\ref{tab:fitres}),
which has a minimal exclusion level of $3.8\sigma$.
The $\dlnl$ value observed in data (4.8)
is indicated with an arrow.}
\label{fig:toy}
\end{figure}

We also generate MC pseudoexperiments in accordance with the fit results for
the $0^{++}$ hypothesis and obtain the distribution of $\dlnl$.
This distribution is fitted to an asymmetric Gaussian function and the
confidence level of the $0^{++}$ hypothesis is calculated as the integral
of the fitting function normalized to 1 from $-\infty$ to the value of
$\dlnl$ in data.
The resulting confidence levels are shown in Table~\ref{tab:excl}.
The distributions of $\dlnl$ for the default model are shown in
Fig.~\ref{fig:toy}.

\begin{table}
\caption{Exclusion levels of the $2^{++}$ hypothesis and confidence levels of
the $0^{++}$ hypothesis. If there are multiple solutions, the results for the
solution with the smallest exclusion level are shown.}
\begin{tabular}{c|c|c}
\hline\hline
Model & $2^{++}$ exclusion & $0^{++}$ CL \\
\hline
Default (constant nonresonant) & $3.8\sigma$ & 77\% \\
NRQCD nonresonant              & $3.4\sigma$ & 73\% \\
$M_{\DD}^{-4}$ nonresonant     & $3.6\sigma$ & 70\% \\
Background mass calculation    & $3.8\sigma$ & 75\% \\
Optimization ($a=4$)           & $3.3\sigma$ & 69\% \\
Optimization ($a=6$)           & $3.4\sigma$ & 64\% \\
Free $r$                       & $2.5\sigma$ & 74\% \\
\hline\hline
\end{tabular}
\label{tab:excl}
\end{table}

\subsection{Cross section}

The cross section calculation is similar to that used in the previous
analyses~\cite{Abe:2007jna,Abe:2007sya}. The full cross section can be
calculated from Eq.~\eqref{eq:nsig} summed over all $\D$
decay channels with the
efficiency corrected by weighting generated and reconstructed events in
accordance with the measured signal PDF. The cross section for the
process $\elp \elm \to \jp \xst{3860} (\to \DD)$ includes the $\xst{3860}$
fit fraction, which is calculated as
$\int S_{\xst{3860}}(\Phi) d\Phi / \int S(\Phi) d\Phi$, where
$S_{\xst{3860}}(\Phi)$ is the signal PDF with the $S_{\xst{3860}}(\Phi)$
contribution only. The cross sections are calculated at all energy points
with results listed in Table~\ref{tab:cross_section}.

The cross section is corrected for differences
between the particle identification requirement efficiencies in
data and MC, which are obtained from a $D^{*+}\to D^0(\to K^-\pi^+)\pi^+$
control sample for $K$ and $\pi$ and a sample of $\gamma\gamma\to\lp\lm$
events for $\mu$ and $e$. The ratio of the particle identification efficiency
in data and MC averaged over all $D$ decay channels is $(92.4 \pm 4.1)\%$.

The cross section is also corrected for differences between the
$\ks$ and $\piz$ reconstruction efficiencies in data and MC. The $\ks$
efficiency correction is determined from a sample of
$\Dz \to \ks \pip \pim$ decays.
The efficiency ratio is $(99.1 \pm 2.3)\%$
 for the channel $\dcha{9}$
and is slightly different for other channels with a daughter $\ks$
because of the different $\ks$ momentum distribution.
The average $\ks$ efficiency ratio is $(99.8 \pm 0.4)\%$.
This ratio, as well as the average $\piz$ efficiency ratio described below,
is closer to 100\% and has a small error
since only some $D$ decay channels have a daughter $\ks$ or $\piz$.

The $\piz$ efficiency correction is determined from a
$\tau^- \to \pim \piz \nu_\tau$ control sample~\cite{Ryu:2014vpc}.
Note that the resulting correction differs from the result of
Ref.~\cite{Ryu:2014vpc} because the $\piz$ momentum distributions of the $\piz$
in individual $\D$ decay channels differ
from the momentum distribution in Ref.~\cite{Ryu:2014vpc}. For example,
the efficiency ratio is $(95.9 \pm 2.2)\%$ for the channel $\dcha{10}$.
The average $\piz$ efficiency ratio in data and MC is $(99.3 \pm 0.4)\%$.

The systematic error sources are listed in Table~\ref{tab:cross_sect_syst}.
Some of these are the same as for the $\xst{3860}$ mass and width.
They include the nonresonant amplitude model dependence,
the optimization variation and the background mass calculation uncertainty.
The uncertainty related to the alternative signal model defined
in Eq.~\eqref{eq:alternative_signal} is not
included since this model does not distinguish the resonant and nonresonant
contributions.
Other systematic error sources include the efficiency difference between
the training and testing signal samples,
uncertainties due to the statistical errors of the signal PDF
parameters, luminosity, track reconstruction, $\jp$ and $D$ branching
fractions and corrections due to the
differences between the efficiency in data and MC for particle identification
requirements and $\ks$ and $\piz$ reconstruction efficiencies.

\begin{table}
\caption{Born cross section measurement results.}
\begin{tabular}{c|c|c}
\hline\hline
Data set & Energy, $\gev$ &
$\sigma_{\elp \elm \to \jp \xst{3860} (\to \DD)}^\mathrm{(Born)}$, $\fb$ \\
\hline
$\Upsilon(1S)$ & 9.46 & $77\err{66}{66}\err{9}{7}$\\
$\Upsilon(2S)$ & 10.02 & $6.9\err{12.6}{12.6}\err{0.9}{0.7}$ \\
$\Upsilon(3S)$ & 10.36 & $77\err{85}{85}\err{11}{8}$ \\
Continuum      & 10.52 & $5.5\err{5.7}{5.7}\err{0.7}{0.5}$ \\
$\Upsilon(4S)$ & 10.58 & $21.7\err{3.9}{4.3}\err{2.9}{2.1}$ \\
$\Upsilon(5S)$ & 10.87 & $17.9\err{7.2}{7.3}\err{2.4}{1.8}$ \\
\hline\hline
\end{tabular}
\label{tab:cross_section}
\end{table}

\begin{table}
\caption{Relative systematic errors of the Born cross section
for $\elp \elm \to \jp \xst{3860} (\to \DD)$.}
\begin{tabular}{c|c}
\hline\hline
Source & Error \\
\hline
Nonresonant amplitude model       & $\err{10.3}{0.0}$\% \\
Optimization                      & $\err{0.0}{5.1}$\% \\
Background mass calculation       & $\err{2.1}{0.0}$\% \\
Training and testing difference   & $<$0.1\% \\
Signal PDF statistical error      & $\err{5.6}{5.8}$\% \\
Luminosity                        & $\pm $1.4\% \\
Tracking                          & $\pm $1.7\% \\
$\jp$ and $D$ branching fractions & $\pm $3.3\% \\
Particle identification           & $\pm $4.3\% \\
$\ks$ reconstruction              & $\pm $0.4\% \\
$\piz$ reconstruction             & $\pm $0.4\% \\
\hline
Total & $\err{13.2}{9.7}$\% \\
\hline\hline
\end{tabular}
\label{tab:cross_sect_syst}
\end{table}

\section{Conclusions and discussion}

In summary, a new charmoniumlike state, the $\xst{3860}$, is observed
in the process $\jpdd$. The $\xst{3860}$ mass is
$(3862\err{26}{32}\err{40}{13})\ \mevcc$ and its width is
$(201\err{154}{67}\err{88}{82})\ \mev$.
The $J^{PC}=0^{++}$ hypothesis is favored over the $2^{++}$ hypothesis at the
level of $2.5\sigma$.


The new state $\xst{3860}$ seems to be a better candidate for the $\chicp$
charmonium state than the $\x{3915}$, since its
properties are well matched to expectations for the $\chicp$ resonance.
The preferred quantum numbers of the $\xst{3860}$ are $J^{PC}=0^{++}$,
although the $2^{++}$ hypothesis is not excluded.
The measured $\xst{3860}$ mass is close to
potential model expectations for the $\chicp$. For example,
the predicted mass in the Ebert-Faustov-Galkin model
is $3854\ \mevcc$~\cite{Ebert:2002pp}. Although the Godfrey-Isgur
model~\cite{Godfrey:1985xj} prediction ($3916\ \mevcc$) is somewhat higher,
the differences between this model's predicted and the observed masses
for the established $\chi_{c2}(2P)$ and $\psi(4040)$ charmonium states are
also high by a similar amount (${\sim}60\ \mevcc$).
Potential models generally predict that the value of the mass-splitting ratio
\begin{equation}
r_c = \frac{m_{\chi_{c2}(2P)} - m_{\chi_{c0}(2P)}}
{m_{\chi_{c2}(1P)} - m_{\chi_{c0}(1P)}}
\end{equation}
lies between 0.6 and 0.9~\cite{Olsen:2014maa};
if the $\xst{3860}$ is indeed the $\chicp$,
then $r_c = 0.46\err{0.25}{0.34}$.
As a conventional charmonium state above the $\DD$
threshold, the $\chicp$ is expected to decay primarily to $\DD$,
which coincides with our observed decay mode of the $\xst{3860}$
[unlike the $X(3915)$, where the $\DD$ mode has not been seen and
the actual $\jp \omega$ observation mode is expected to be OZI-suppressed for
the $\chicp$].
The $\xst{3860}$ helicity amplitudes are consistent with pure $S$-wave
production.
An angular analysis of the related process $\elp \elm \to \jp \chi_{c0}(1P)$
was performed in Ref.~\cite{Abe:2004ww}; the angular distribution of this
process was also found to be consistent with pure $S$-wave production.
Note that while the production amplitudes for the $\xst{3860}$ and
$\chi_{c0}(1P)$ are in mutual agreement, they do not agree with the NRQCD
prediction~\cite{Liu:2004ga}.
In addition, the $\xst{3860}$ mass and width agree
with the $\chicp$ parameters determined from an alternative fit to
the Belle and \babar $\gamma \gamma \to \DD$ data performed in
Ref.~\cite{Guo:2012tv}:
$M = (3837.6 \pm 11.5)\ \mevcc$, $\Gamma = (221 \pm 19)\ \mev$.

\section*{Acknowledgment}

We thank the KEKB group for the excellent operation of the
accelerator; the KEK cryogenics group for the efficient
operation of the solenoid; and the KEK computer group,
the National Institute of Informatics, and the 
PNNL/EMSL computing group for valuable computing
and SINET5 network support.  We acknowledge support from
the Ministry of Education, Culture, Sports, Science, and
Technology (MEXT) of Japan, the Japan Society for the 
Promotion of Science (JSPS), and the Tau-Lepton Physics 
Research Center of Nagoya University; 
the Australian Research Council;
Austrian Science Fund under Grant No.~P 26794-N20;
the National Natural Science Foundation of China under Contracts 
No.~10575109, No.~10775142, No.~10875115, No.~11175187, No.~11475187, 
No.~11521505 and No.~11575017;
the Chinese Academy of Science Center for Excellence in Particle Physics; 
the Ministry of Education, Youth and Sports of the Czech
Republic under Contract No.~LTT17020;
the Carl Zeiss Foundation, the Deutsche Forschungsgemeinschaft, the
Excellence Cluster Universe, and the VolkswagenStiftung;
the Department of Science and Technology of India; 
the Istituto Nazionale di Fisica Nucleare of Italy; 
the WCU program of the Ministry of Education, National Research Foundation (NRF)
of Korea Grants No.~2011-0029457, No.~2012-0008143,
No.~2014R1A2A2A01005286,
No.~2014R1A2A2A01002734, No.~2015R1A2A2A01003280,
No.~2015H1A2A1033649, No.~2016R1D1A1B01010135, No.~2016K1A3A7A09005603, No.~2016K1A3A7A09005604, No.~2016R1D1A1B02012900,
No.~2016K1A3A7A09005606, No.~NRF-2013K1A3A7A06056592;
Korean Institute for Basic Science Project Code IBS-R016-D1;
the Brain Korea 21-Plus program, Radiation Science Research Institute, Foreign Large-size Research Facility Application Supporting project and the Global Science Experimental Data Hub Center of the Korea Institute of Science and Technology Information;
the Polish Ministry of Science and Higher Education and 
the National Science Center;
the Ministry of Education and Science of the Russian Federation
under Contracts No.~14.A12.31.0006 and No.~3.2989.2017 and
the Russian Foundation for Basic Research;
the Slovenian Research Agency;
Ikerbasque, Basque Foundation for Science and
the Euskal Herriko Unibertsitatea (UPV/EHU) under program UFI 11/55 (Spain);
the Swiss National Science Foundation; 
the Ministry of Education and the Ministry of Science and Technology of Taiwan;
and the U.S.\ Department of Energy and the National Science Foundation.


\appendix
\section{Derivation of the signal density function}
\label{sec:amplitude}

\begin{figure}[ht]
\includegraphics[width=6cm]{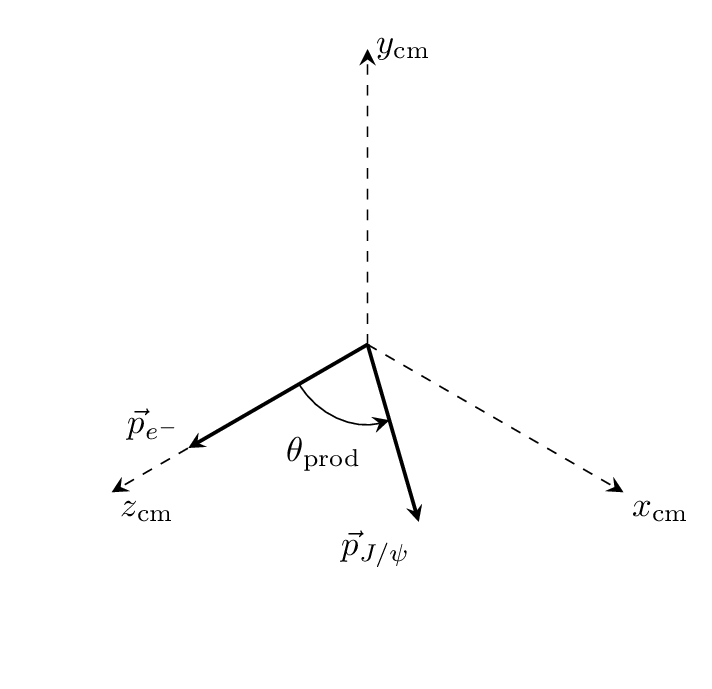}
\caption{Definition of the production angle (in the $\elp \elm$
center-of-mass frame).}
\label{fig:thetaproduction}
\end{figure}

The signal density function is calculated for the process
$\elp \elm \to \jp \Xst$,
with $\jp \to \lp \lm$ and $\Xst \to \DD$.
The definition of the production angle $\thprod$ is shown
in Fig.~\ref{fig:thetaproduction}.
This angle is given by
\begin{equation}
\cos \thprod = \frac{\vec{p}_\elm \cdot \vec{p}_\jp}
{|\vec{p}_\elm| |\vec{p}_\jp|},
\end{equation}
where $\vec{p}_\elm$ and $\vec{p}_\jp$ are the momenta of the beam
$\elm$ and $\jp$, respectively, in the center-of-mass frame.

\begin{figure}[ht]
\includegraphics[width=6cm]{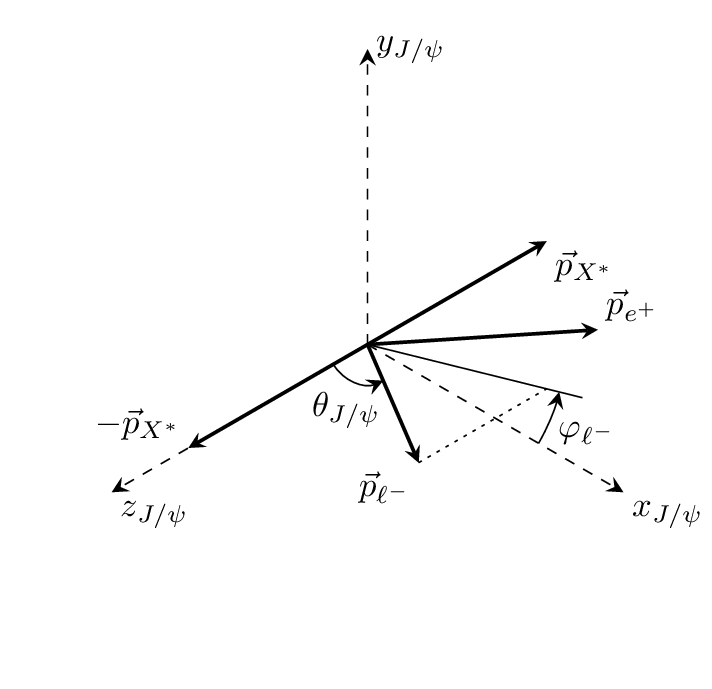}
\caption{Definitions of the $\jp$ helicity angle and
the angle $\varphi_\lm$ (in the $\jp$ rest frame).}
\label{fig:thetajpsi}
\end{figure}

The definitions of the $\jp$ helicity
angle and the $\lm$ azimuthal angle are shown in Fig.~\ref{fig:thetajpsi}.
The $\jp$ helicity angle is given by
\begin{equation}
\cos \theta_\jp = -\frac{{\vec{p}_\Xst}\cdot{\vec{p}_\lm}}
{|\vec{p}_\Xst||\vec{p}_\lm|},
\end{equation}
where $\vec{p}_\Xst$ and $\vec{p}_\lm$ are
the momenta of $\Xst$ and $\lm$, respectively, in the $\jp$ rest frame.
The azimuthal angle $\varphi_\lm$ can be expressed as
\begin{equation}
\begin{aligned}
& \cos \varphi_\lm = \frac{{\vec{a}_\elp}\cdot{\vec{a}_\lm}}
{|\vec{a}_\elp||\vec{a}_\lm|}, \\
& \sin \varphi_\lm = -\frac{[\vec{p}_\Xst\times\vec{a}_\elp]
\cdot{\vec{a}_\lm}} {|\vec{p}_\Xst||\vec{a}_\elp||\vec{a}_\lm|}, \\
\end{aligned}
\end{equation}
where
\begin{equation}
\begin{aligned}
& \vec{a}_\elp = \vec{p}_\elp - \frac{{\vec{p}_\elp}\cdot{\vec{p}_\Xst}}
{|\vec{p}_\Xst|^2} \vec{p}_\Xst, \\
& \vec{a}_\lm = \vec{p}_\lm - \frac{{\vec{p}_\lm}\cdot{\vec{p}_\Xst}}
{|\vec{p}_\Xst|^2} \vec{p}_\Xst, \\
\end{aligned}
\end{equation}
with $\vec{p}_\elp$ denoting the momentum of a beam positron in the $\jp$ rest
frame.

\begin{figure}[ht]
\includegraphics[width=6cm]{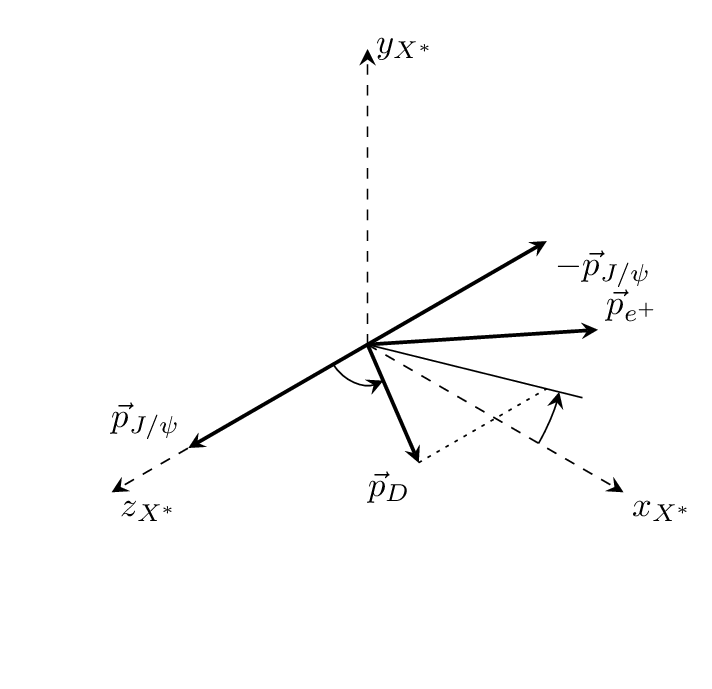}
\caption{The coordinate system $(x_\Xst,y_\Xst,z_\Xst)$ (in the $\Xst$ rest
frame).}
\label{fig:thetax}
\end{figure}

The coordinate system $(x_\mathrm{cm},y_\mathrm{cm},z_\mathrm{cm})$ is
rotated by $\thprod$ around the $y$ axis and boosted to the $\jp$ or $\Xst$ rest
frame. For the $\jp$, the resulting coordinate system $(x_\jp,y_\jp,z_\jp)$
has normal orientation: the $\jp$ helicity quantization axis is the same
as $z_\jp$. For the $\Xst$,
the resulting coordinate system $(x_\Xst,y_\Xst,z_\Xst)$ has inverse
orientation: the helicity quantization axis is antiparallel to the $z_\Xst$
axis. The latter coordinate system is shown in Fig.~\ref{fig:thetax}.
One can perform a rotation to the $\D$ helicity frame directly from
the coordinate system $(x_\Xst,y_\Xst,z_\Xst)$; however, this results in
a nonstandard definition of the helicity and azimuthal angles as well as a
sign flip for the $\Xst$ helicity in the amplitude. To
conform to standard definitions, the coordinate system $(x_\Xst,y_\Xst,z_\Xst)$
is rotated by $\pi$ around the $z$ axis and by $\pi$ around the $y$ axis.
The direction of the $z'_\Xst$ axis of the resulting coordinate system
$(x'_\Xst,y'_\Xst,z'_\Xst)$ is opposite the direction of the $z_\Xst$;
the direction of the $x$ axis is the same. The amplitude corresponding to
this rotation is
\begin{equation}
\begin{aligned}
A_{\lambda_\Xst\,\tilde{\lambda}_\Xst}
 & = \Dfun{J(\Xst)\,*}{-\lambda_\Xst}{\lambda_\Xst}{\pi}{\pi}{0} \\
& = e^{-i \lambda_\Xst \pi}
\dfun{J(\Xst)}{-\lambda_\Xst}{\lambda_\Xst}{\pi} = (-1)^{J(\Xst)},
\end{aligned}
\end{equation}
where $\lambda_\Xst$ is the $\Xst$ helicity.
Thus, the amplitude of the additional rotation merely introduces a common term
for all $\Xst$ helicity amplitudes and can be omitted.
Note that the amplitude of the process
$B^+ \to X (\to \jp (\to \mup \mum) \phi (\to \kp \km)) \kp$
in Ref.~\cite{Aaij:2016nsc}
is defined in a similar way without the additional term $(-1)^{J(\phi)}$.

\begin{figure}[ht]
\includegraphics[width=6cm]{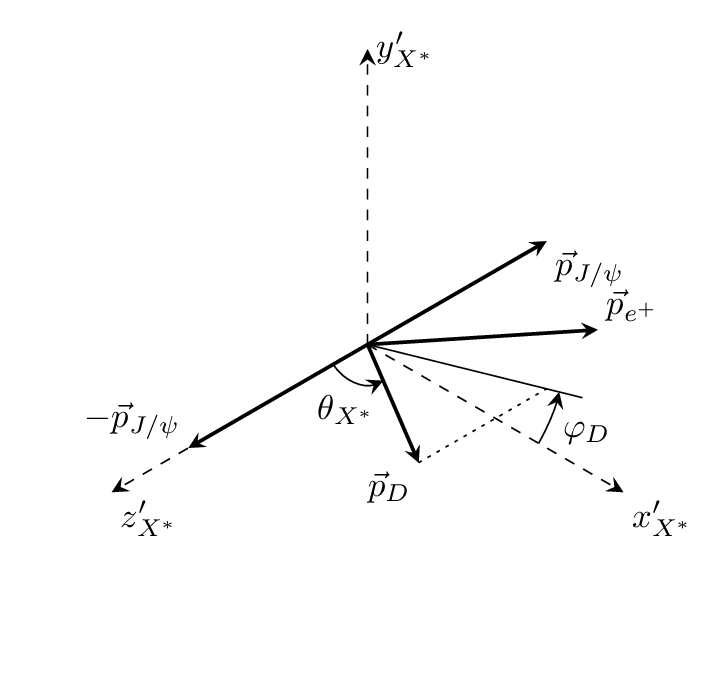}
\caption{Definitions of the $\Xst$ helicity angle and
the angle $\varphi_\D$ (in the $\Xst$ rest frame).}
\label{fig:thetax2}
\end{figure}

The definitions of the $\Xst$ helicity
angle and the $\D$ azimuthal angle are shown in Fig.~\ref{fig:thetax2}.
The $\Xst$ helicity angle is given by
\begin{equation}
\cos \theta_\Xst = -\frac{{\vec{p}_\jp}\cdot{\vec{p}_D}}
{|\vec{p}_\jp||\vec{p}_D|},
\end{equation}
where $\vec{p}_\jp$ and $\vec{p}_D$ are
the momenta of $\jp$ and $D$, respectively, in the $\Xst$ rest frame.
Note that $\vec{p}_\D$ is the momentum of the reconstructed $D$ meson;
it changes to $\vec{p}_\Db$ if the $\Db$ meson is reconstructed.
The azimuthal angle $\varphi_\D$ can be expressed as
\begin{equation}
\begin{aligned}
& \cos \varphi_\D = \frac{{\vec{a}_\elp}\cdot{\vec{a}_\D}}
{|\vec{a}_\elp||\vec{a}_\D|}, \\
& \sin \varphi_\D = -\frac{[\vec{p}_\jp\times\vec{a}_\elp]\cdot{\vec{a}_\D}}
{|\vec{p}_\jp||\vec{a}_\elp||\vec{a}_\D|}. \\
\end{aligned}
\end{equation}
where
\begin{equation}
\begin{aligned}
& \vec{a}_\elp = \vec{p}_\elp - \frac{{\vec{p}_\elp}\cdot{\vec{p}_\jp}}
{|\vec{p}_\jp|^2} \vec{p}_\jp, \\
& \vec{a}_\D = \vec{p}_\D - \frac{{\vec{p}_\D}\cdot{\vec{p}_\jp}}
{|\vec{p}_\jp|^2} \vec{p}_\jp, \\
\end{aligned}
\end{equation}
with $\vec{p}_\elp$ denoting the momentum of a beam positron
in the $\jp$ rest frame.

The amplitude is given by
\begin{equation}
\begin{aligned}
& A_{\lbeam\,\lambda_\jp\,\lambda_\Xst\,\lambda_\leplep}(\Phi) \\
& \qquad =
  H_{\lambda_\jp\,\lambda_\Xst}
  \dfun{1}{\lbeam}{\lambda_\jp-\lambda_\Xst}{\thprod}
  e^{i \lambda_\jp \varphi_\lm} \\
& \qquad \qquad \times
  \dfun{1}{\lambda_\jp}{\lambda_\leplep}{\theta_\jp}
  e^{i \lambda_\Xst \varphi_\D}
  \dfun{J(\Xst)}{\lambda_\Xst}{0}{\theta_\Xst}, \\
\end{aligned}
\end{equation}
and
\begin{equation}
A_{\lbeam\,\lambda_\leplep}(\Phi) =
\kern-20pt\sum\limits_{\substack{\lambda_\jp = -1, 0, 1 \\
\lambda_\Xst = -J(\Xst), ..., J(\Xst)}}\kern-20pt
A_{\lbeam\,\lambda_\jp\,\lambda_\Xst\,\lambda_\leplep}(\Phi)
\end{equation}
where $H_{\lambda_\jp\,\lambda_\Xst}$ is the production helicity
amplitude,
$\lambda$ is the helicity of the particle specified by the index,
$\lbeam$ is the sum of helicities of the beam electron and positron and
$\lambda_\leplep$ is the sum of the helicities of the $\jp$ decay products.

The $\Xst$ spin and parity are related:
\begin{equation}
P(\Xst) = (-1)^{J(\Xst)}.
\label{eq:xjp}
\end{equation}
The helicity amplitudes are related due to parity conservation; taking into
account Eq.~\eqref{eq:xjp}, the relation is
\begin{equation}
H_{-\lambda_\jp\,-\lambda_\Xst} = H_{\lambda_\jp\,\lambda_\Xst},
\end{equation}
independent of $J(\Xst)$. An additional selection rule is
\begin{equation}
|\lambda_\jp-\lambda_\Xst| \leq 1.
\end{equation}
The lepton pair is produced in the electromagnetic decay $\jp\to\lp\lm$
via a virtual photon; thus, its helicity $\lambda_\leplep$ may be equal to
1 or $-1$. In the case of one-photon exchange (for positive $C$-parity of $X$),
the beam helicity is also equal to 1 or $-1$.

The amplitude is additionally constrained due to $C$-parity conservation.
The $\DD$ system has $C = (-1)^{J(\Xst)}$; as its $C$-parity should be
positive in case of one-photon production, odd values of $J(\Xst)$ are
forbidden. Thus, only two hypotheses, $J(\Xst) = 0$ and $J(\Xst) = 2$, are
considered.

The signal density function is
\begin{equation}
S(\Phi) = \kern-10pt\sum\limits_{\substack{\lbeam = -1, 1\\\lambda_\leplep = -1, 1}}
\Big| \sum\limits_\Xst A_{\lbeam\,\lambda_\leplep}(\Phi) A_\Xst(M_\DD)
\Big|^2,
\end{equation}
where $\Xst$ is any contribution to the signal ($\Xst$ resonance or
nonresonant amplitude).
If the reconstructed particle is $\Db$ instead of $\D$, the
angles $\theta_\Xst$ and $\varphi_\D$ can be calculated from the angles
for the reconstructed $\Db$: $\theta_\Xst \to \pi - \theta_\Xst$,
$\varphi_\D \to \varphi_\D + \pi$ (see Fig.~\ref{fig:thetax2}). However,
it is not necessary to perform this substitution, as
the resulting signal density is exactly the same.

For $J(\Xst) = 0$, there are two helicity amplitudes: $H_{0\,0}$ and $H_{1\,0}$.
For $J(\Xst) = 2$, there are five helicity amplitudes: $H_{0\,0}$, $H_{1\,0}$,
$H_{0\,1}$, $H_{1\,1}$ and $H_{1\,2}$.
The helicity amplitudes can be expressed in terms of partial wave amplitudes
$L(\jp,\Xst)_{J(\jp,\Xst)}$ using Eq.~(B5) from
Ref.~\cite{Jacob:1959at}.
For $J(\Xst) = 0$, the relations are
(partial wave amplitudes forbidden by parity conservation being omitted):
\begin{equation}
\begin{aligned}
& H_{0\,0} = \sqrt{\frac{1}{3}} S_{1} - \sqrt{\frac{2}{3}} D_{1}, \\
& H_{1\,0} = \sqrt{\frac{1}{3}} S_{1} + \sqrt{\frac{1}{6}} D_{1}. \\
\end{aligned}
\label{eq:helpwa}
\end{equation}


\end{document}

%% file: authors.tex
\noaffiliation
\affiliation{University of the Basque Country UPV/EHU, 48080 Bilbao}
\affiliation{Beihang University, Beijing 100191}
\affiliation{Budker Institute of Nuclear Physics SB RAS, Novosibirsk 630090}
\affiliation{Faculty of Mathematics and Physics, Charles University, 121 16 Prague}
\affiliation{Chonnam National University, Kwangju 660-701}
\affiliation{University of Cincinnati, Cincinnati, Ohio 45221}
\affiliation{Deutsches Elektronen--Synchrotron, 22607 Hamburg}
\affiliation{University of Florida, Gainesville, Florida 32611}
\affiliation{Justus-Liebig-Universit\"at Gie\ss{}en, 35392 Gie\ss{}en}
\affiliation{SOKENDAI (The Graduate University for Advanced Studies), Hayama 240-0193}
\affiliation{Hanyang University, Seoul 133-791}
\affiliation{University of Hawaii, Honolulu, Hawaii 96822}
\affiliation{High Energy Accelerator Research Organization (KEK), Tsukuba 305-0801}
\affiliation{J-PARC Branch, KEK Theory Center, High Energy Accelerator Research Organization (KEK), Tsukuba 305-0801}
\affiliation{IKERBASQUE, Basque Foundation for Science, 48013 Bilbao}
\affiliation{Indian Institute of Science Education and Research Mohali, SAS Nagar, 140306}
\affiliation{Indian Institute of Technology Bhubaneswar, Satya Nagar 751007}
\affiliation{Indian Institute of Technology Guwahati, Assam 781039}
\affiliation{Indian Institute of Technology Madras, Chennai 600036}
\affiliation{Indiana University, Bloomington, Indiana 47408}
\affiliation{Institute of High Energy Physics, Chinese Academy of Sciences, Beijing 100049}
\affiliation{Institute of High Energy Physics, Vienna 1050}
\affiliation{Institute for High Energy Physics, Protvino 142281}
\affiliation{INFN - Sezione di Napoli, 80126 Napoli}
\affiliation{INFN - Sezione di Torino, 10125 Torino}
\affiliation{Advanced Science Research Center, Japan Atomic Energy Agency, Naka 319-1195}
\affiliation{J. Stefan Institute, 1000 Ljubljana}
\affiliation{Kanagawa University, Yokohama 221-8686}
\affiliation{Center for Underground Physics, Institute for Basic Science, Daejeon 34047}
\affiliation{Institut f\"ur Experimentelle Kernphysik, Karlsruher Institut f\"ur Technologie, 76131 Karlsruhe}
\affiliation{Kennesaw State University, Kennesaw, Georgia 30144}
\affiliation{King Abdulaziz City for Science and Technology, Riyadh 11442}
\affiliation{Department of Physics, Faculty of Science, King Abdulaziz University, Jeddah 21589}
\affiliation{Korea Institute of Science and Technology Information, Daejeon 305-806}
\affiliation{Korea University, Seoul 136-713}
\affiliation{Kyoto University, Kyoto 606-8502}
\affiliation{Kyungpook National University, Daegu 702-701}
\affiliation{\'Ecole Polytechnique F\'ed\'erale de Lausanne (EPFL), Lausanne 1015}
\affiliation{P.N. Lebedev Physical Institute of the Russian Academy of Sciences, Moscow 119991}
\affiliation{Faculty of Mathematics and Physics, University of Ljubljana, 1000 Ljubljana}
\affiliation{Ludwig Maximilians University, 80539 Munich}
\affiliation{University of Malaya, 50603 Kuala Lumpur}
\affiliation{University of Maribor, 2000 Maribor}
\affiliation{Max-Planck-Institut f\"ur Physik, 80805 M\"unchen}
\affiliation{School of Physics, University of Melbourne, Victoria 3010}
\affiliation{University of Miyazaki, Miyazaki 889-2192}
\affiliation{Moscow Physical Engineering Institute, Moscow 115409}
\affiliation{Moscow Institute of Physics and Technology, Moscow Region 141700}
\affiliation{Graduate School of Science, Nagoya University, Nagoya 464-8602}
\affiliation{National Central University, Chung-li 32054}
\affiliation{National United University, Miao Li 36003}
\affiliation{Department of Physics, National Taiwan University, Taipei 10617}
\affiliation{H. Niewodniczanski Institute of Nuclear Physics, Krakow 31-342}
\affiliation{Nippon Dental University, Niigata 951-8580}
\affiliation{Niigata University, Niigata 950-2181}
\affiliation{Novosibirsk State University, Novosibirsk 630090}
\affiliation{Osaka City University, Osaka 558-8585}
\affiliation{Pacific Northwest National Laboratory, Richland, Washington 99352}
\affiliation{University of Pittsburgh, Pittsburgh, Pennsylvania 15260}
\affiliation{Theoretical Research Division, Nishina Center, RIKEN, Saitama 351-0198}
\affiliation{University of Science and Technology of China, Hefei 230026}
\affiliation{Showa Pharmaceutical University, Tokyo 194-8543}
\affiliation{Soongsil University, Seoul 156-743}
\affiliation{Stefan Meyer Institute for Subatomic Physics, Vienna 1090}
\affiliation{Sungkyunkwan University, Suwon 440-746}
\affiliation{School of Physics, University of Sydney, New South Wales 2006}
\affiliation{Department of Physics, Faculty of Science, University of Tabuk, Tabuk 71451}
\affiliation{Tata Institute of Fundamental Research, Mumbai 400005}
\affiliation{Department of Physics, Technische Universit\"at M\"unchen, 85748 Garching}
\affiliation{Toho University, Funabashi 274-8510}
\affiliation{Department of Physics, Tohoku University, Sendai 980-8578}
\affiliation{Earthquake Research Institute, University of Tokyo, Tokyo 113-0032}
\affiliation{Department of Physics, University of Tokyo, Tokyo 113-0033}
\affiliation{Tokyo Institute of Technology, Tokyo 152-8550}
\affiliation{Tokyo Metropolitan University, Tokyo 192-0397}
\affiliation{University of Torino, 10124 Torino}
\affiliation{Virginia Polytechnic Institute and State University, Blacksburg, Virginia 24061}
\affiliation{Wayne State University, Detroit, Michigan 48202}
\affiliation{Yamagata University, Yamagata 990-8560}
\affiliation{Yonsei University, Seoul 120-749}
  \author{K.~Chilikin}\affiliation{P.N. Lebedev Physical Institute of the Russian Academy of Sciences, Moscow 119991}\affiliation{Moscow Physical Engineering Institute, Moscow 115409} 
  \author{I.~Adachi}\affiliation{High Energy Accelerator Research Organization (KEK), Tsukuba 305-0801}\affiliation{SOKENDAI (The Graduate University for Advanced Studies), Hayama 240-0193} 
  \author{H.~Aihara}\affiliation{Department of Physics, University of Tokyo, Tokyo 113-0033} 
  \author{S.~Al~Said}\affiliation{Department of Physics, Faculty of Science, University of Tabuk, Tabuk 71451}\affiliation{Department of Physics, Faculty of Science, King Abdulaziz University, Jeddah 21589} 
  \author{D.~M.~Asner}\affiliation{Pacific Northwest National Laboratory, Richland, Washington 99352} 
  \author{V.~Aulchenko}\affiliation{Budker Institute of Nuclear Physics SB RAS, Novosibirsk 630090}\affiliation{Novosibirsk State University, Novosibirsk 630090} 
  \author{R.~Ayad}\affiliation{Department of Physics, Faculty of Science, University of Tabuk, Tabuk 71451} 
  \author{V.~Babu}\affiliation{Tata Institute of Fundamental Research, Mumbai 400005} 
  \author{I.~Badhrees}\affiliation{Department of Physics, Faculty of Science, University of Tabuk, Tabuk 71451}\affiliation{King Abdulaziz City for Science and Technology, Riyadh 11442} 
  \author{A.~M.~Bakich}\affiliation{School of Physics, University of Sydney, New South Wales 2006} 
% ...
\collaboration{The Belle Collaboration}